\documentclass[useAMS,usegraphicx,usenatbib]{mn2e}
\usepackage{amsmath} 
\usepackage{subeqn, color,amssymb}

\def\be{\begin{equation}}
\def\ee{\end{equation}}

\def\deg{\circ}

\def\epssf{\epsilon_{\rm SF}}

\def\msun{M_{\odot}}

\title[Galactic Centre Warped Discs]
{On the possibility of a warped disc origin of the inclined stellar discs at the Galactic Centre}
\author[A. Ulubay-Siddiki, H. Bartko \& O. Gerhard]
{A. Ulubay-Siddiki$^{1}$\thanks{E-mail: aulubays@istanbul.edu.tr, hbartko@mpe.mpg.de, gerhard@mpe.mpg.de}, 
H. Bartko$^{2}$, and O. Gerhard$^{2}$\\
$^{1}$Istanbul University, Faculty of Science, Department of Physics, 34134 Vezneciler, Istanbul, Turkey\\
$^{2}$Max-Planck-Institut f\"ur Extraterrestrische Physik,Giessenbachstra{\ss}e, 
D-85748 Garching, Germany}

\begin{document}
\maketitle
\label{firstpage}
\begin{abstract}

The central parsec of our Galaxy hosts a population of young stars. 
At distances of $r \sim$ 0.03 to 0.5 pc, most of these stars seem to 
form a system of mutually inclined discs of clockwise and counter clockwise 
rotating stars. We present a possible warped disc origin scenario for these 
stars assuming that an initially flat accretion disc becomes warped 
due to a central radiation source via the Pringle instability, or due to a 
spinning black hole via the Bardeen-Petterson effect, before it cools, 
fragments, and forms stars. From simple arguments 
we show that this is plausible if the star formation efficiency 
is high, $\epsilon_{\rm SF} \lesssim 1$, and the viscosity 
parameter $\alpha \sim 0.1$. After fragmentation, we model the 
disc as a collection of concentric, circular rings tilted with 
respect to each other, and construct time evolution models of warped 
discs for mass ratios and other parameters relevant to the Galactic 
Centre environment, but for also more massive discs.
We take into account the disc's self-gravity in the non-linear 
regime and the torques 
exerted by a slightly flattened surrounding star cluster. 
Our simulations show that a self-gravitating low-mass disc ($ M_{\rm d} /M_{\rm bh} \sim 0.001$) 
precesses with its integrity maintained in the
life-time of the stars, but precesses essentially freely when the
torques from a non-spherical cluster are included. An
intermediate-mass disc ($ M_{\rm d} /M_{\rm bh} \sim 0.01$) breaks into pieces which precess
as independent discs in the self-gravity-only case, and become
disrupted in the presence of the star cluster torques. Finally, for a
high mass disc ($ M_{\rm d} /M_{\rm bh} \sim 0.1$) the evolution is dominated by self-gravity
and the disc is broken but not dissolved. The time-scale after which
the disc breaks into pieces scales almost linearly with $M_{\rm d} /M_{\rm bh}$ for
self-gravitating models. Typical values are longer than the age of the
stars for $ M_{\rm d} /M_{\rm bh} \sim 0.001$, and are in the range 
$\sim 8 \times 10^4-10^5$ yr for $ M_{\rm d} /M_{\rm bh} \sim 0.1-0.01$ respectively. 
None of these discs explain the two Galactic Centre discs with their rotation
properties. A comparison of the models with the
better-defined clockwise rotating disc shows that the lowest mass
model in a spherical star cluster matches the data best.
\end{abstract}

\section{Introduction}
 \label{sec:introchap2} 
The centre of our Galaxy hosts a supermassive black hole (SMBH), SgrA*, with a 
mass of $3.95 \pm 0.06 \times 10^6 \ M_{\odot}$ 
\citep{genzel00,ghez05, trippe08, gillessen09}. SgrA* is surrounded by a 
cluster of old \citep{trippe08, schoedel09}, as well as a 
group of young stars \citep{krabbe95, genzel03, levin03, paumard06, lu06, lu09, 
bartko09, bartko010}.
 
Of the 136 young stars observed at distances $\sim 0.05$ pc to $\sim 0.5$ pc, 
59 populate a disc \citep{genzel03, levin03, paumard06,lu06,lu09, bartko010} 
which is observed to rotate clockwise (CW) on the plane of sky. Of the rest, 20 
stars seem to populate an other disc highly inclined to the CW one 
\citep{genzel03, paumard06,bartko09, bartko010}, and rotating counter 
clockwise (CCW) on the sky (but also see \cite{lu06}). 
Ages of these young stars are consistent with being a few Myr, suggesting that
there has been a star formation episode in the Galactic Centre (GC) a few million 
years ago.
 
In order for a molecular cloud near a SMBH to fragment into stars, its self-gravity 
should overcome the tidal field of the black hole. This requirement for star 
formation poses a constraint on the minimum cloud densities, which are orders of 
magnitude higher than the observed cloud densities near the 
GC. However theoretical estimates suggest that the 
fragmentation conditions are met naturally on the accretion discs which become 
self-gravitating beyond a few tenth of parsec \citep{kolykhalov80,gammie01,goodman03}. 
Therefore, several numerical simulations have been performed aiming at modeling the 
in-situ fragmentation of a nuclear/accretion disc for parameters relevant to the 
GC. The simulations were run either assuming $a \ priori$  
gravitationally unstable accretion disc already in place 
\citep{nayakshin06,alexander08}, or trying to account 
also for the formation of the disc itself through infall of molecular clouds 
into the vicinity of the black hole \citep{bonnell08, mapelli08, hobbs09}. 

Today it looks like a star forming disc at the GC can be simulated, albeit 
perhaps for a somewhat fine-tuned parameter range. On the other hand, apart from 
the problem of  youth, another issue still to be addressed is the distribution 
of the inclinations of the stars. 
It is reasonable to expect that a planar accretion disc leaves behind a planar 
distribution of stars when it fragments, but the recent data published by \cite{bartko09} 
provide evidence for a warp in the CW disc with an amplitude of about 
$60^{\deg}$. \cite{lu09} point out that even though the 
stars might have formed in-situ, their current
orbital distribution suggests a more sophisticated origin 
than a simple thin accretion disc. Simulations performed 
by \citet{cuadra08} are in line with this idea showing that once 
the stars form on a cold accretion disc it is not possible to 
perturb these stars to the high inclinations at which they 
are observed. \citet{kocsis011} showed that vector resonant 
relaxation between the disc and the surrounding old star cluster might 
excite a warp in the disc if the stochastic torques from the star 
cluster dominate over the disc's self-gravity. 
Recently, \citet{haas010} considered the evolution of the stellar disc 
using N-body simulations. Their simulations successfully produced the 
relative inclination between the stellar disc and the surrounding 
Circumnuclear Disc, however the origin of the disc was not addressed.

Warped discs, although monitored only through maser emission from gas discs, 
exist on similar scales in other nearby galactic nuclei such as 
NGC4258 \citep{herrnstein}, NGC1068 \citep{greenhill97}, and 
Circinus \citep{greenhill03}. An initially planar accretion disc could 
become warped when torqued by a spinning black hole 
\citep{bardeen75, armitage99, lodato07, martin08}, when exposed to 
radiation from a central source \citep{petterson77, pringle96, pringle97}, 
or when subject to the torques induced by the stars in 
a stellar cusp around the black hole \citep{bregman09}.
\cite{milo04} pointed out that the maser nuclei, and the 
Galactic Centre might represent different epochs of a cycle 
during the lifetime of a typical spiral galaxy.

In this paper, we investigate alternative scenarios for the 
formation of a star-forming, warped disc at the Galactic Centre. 
Accordingly, a flat accretion disc forms around SgrA*, extending out
to the location of the young stars observed today. During a supposed
period of active accretion, the disc is illuminated by the central
source, or torqued by a spinning black hole, and becomes warped due to
the Pringle instability or the Bardeen-Petterson effect,
respectively. When the AGN activity subsides, the disc cools and forms
stars. Afterwards, the stellar disc evolves in the gravitational field 
of the black hole, its own self-gravity, and the surrounding old star 
cluster. 
We investigate under which conditions this scenario
could work, and show that with a low mass such as inferred today the
remnant warped stellar disc largely survives for the 
life-time of the observed young stars.

In \S \ref{sec:warpingGC} we work out the conditions for which 
the disc would become warped due to the mechanisms mentioned above, 
and then fragment into stars. In \S \ref{sec:model} we describe 
our model and the numerical scheme for studying the subsequent 
time evolution of the warped stellar disc. The results of our 
simulations are presented in \S \ref{sec:results} where a comparison 
with the observations is also made. In \S \ref{sec:discussion} we 
present a discussion, and in \S \ref{sec:summary} we 
summarize our results, and our conclusions.
\section{Warping the Galactic Centre Disc}
\label{sec:warpingGC}
 In this section, we discuss a plausible scenario of how a disc of
young stars in the Galactic Centre might have acquired its warped
shape. 
We start with the assumption that an accretion disc builds
up, leading to an active phase of sub-Eddington accretion onto the
Galactic Centre black hole. We then investigate two possible
mechanisms: radiation pressure instability 
\citep{petterson77,pringle96,pringle97} 
and Bardeen-Petterson effect \citep{bardeen75} for warping the 
accretion disc, which have both been
extensively discussed in the context of the maser discs in nearby 
Seyfert galaxies 
\citep{maloney96, scheuer96, pringle97, armitage99, lodato07, martin08}. 

We further assume that after some time the accretion and energy production is reduced,
which is followed by a period where the warped disc can cool and form stars. Thereafter
the stellar disc is only subject to gravity, precessing under
the influence of the gravitational torques from the disc itself, and the surrounding 
old star cluster.

In \S \ref{sec:properties}, we constrain the surface density of the disc prior
to fragmentation from the observed number density of young stars.  In
the following subsections \ref{sec:radwarp} and \ref{sec:bardeen}, 
we consider in turn warping by the radiation pressure instability 
and the Bardeen-Petterson effect. In \S \ref{sec:comparetrq}, 
we compare the radiation, viscous, and
gravitational torques on the disc, and in \S \ref{sec:interplay}
we consider fragmentation and star formation.

\subsection{Surface Density of the Disc Prior to Fragmentation}
\label{sec:properties}
The warping mechanisms which we will briefly consider in 
sections (\ref{sec:radwarp}) and (\ref{sec:bardeen}) are generally studied 
within the framework of viscous, steady-state 
accretion discs. As such they make use of a number of  parameters which 
from the observations of the stellar discs can not be tested or 
constrained. For these, we will mostly refer to the canonical values for 
AGN discs when needed. Still, there is one parameter, the surface density 
of the supposed gaseous disc which may be determined by the observations. 
To do so  we make use of the stellar number counts on the GC discs. The total number 
of stars $N_{\star}$ in a stellar population can be calculated by writing
\begin{equation}
N_{\star} = \int_{M1}^{M2} \xi(M) dM,
\label{eq:nstar}
\end{equation}
where $\xi(M)$ is the initial mass function, IMF,  
and $M_1$ and $M_2$ are the 
lowest and highest stellar masses assumed to exist in this population. The 
IMF describes how the 
mass is distributed in stars in a newly born population. For a given IMF, the total 
mass in stars, $M_s$, is calculated from
\begin{equation}
M_{\rm s} = \int_{M1}^{M2}M \xi(M)dM.
\label{eq:mstar}
\end{equation}
 \cite{bartko010} deduce an IMF of $\xi(M)=\xi_{0} M^{-0.45}$ for the GC discs. 
The \cite{bartko010} sample includes 59 stars in the CW disc, 
and 20 stars in the CCW disc. Using equations 
(\ref{eq:nstar}) and (\ref{eq:mstar}), 
and assuming a lower mass end of $1 M_{\odot}$, and an upper mass end of 
$120 M_{\odot}$, the current total stellar mass in the discs can be found to be 
$ M_{\rm s}|_{\rm CW}+ M_{\rm s}|_{\rm CCW} =  M_{\rm s} \sim 5360 M_{\odot}$, where 
$ M_{\rm s}|_{\rm CW}$, and  $ M_{\rm s}|_{\rm CCW}$ are the masses of the CW and CCW 
rotating discs respectively. The mass of the Galactic central black hole is 
$\sim 4 \times 10^6 M_{\rm \odot}$, so the inferred disc mass 
corresponds to a mass ratio $M_{\rm d}/M_{\rm bh}=0.00134$.

\begin{figure}
\begin{centering}
\includegraphics[width=8.5cm]{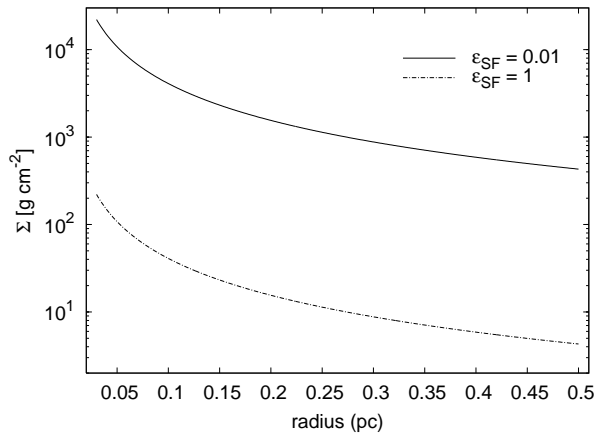}
\caption[Surface density of the GC gaseous disc deduced from the IMF]
{Surface density of the supposed GC gaseous disc for star formation 
efficiencies of $0.01$ (solid line), and $1$ (dotted line).}
\label{fig:sigmaimf}
\end{centering}
\end{figure}

The total stellar disc mass inferred today is 
a fraction of the mass of the original gaseous disc since presumably 
not all the gas was converted to stars. 
Assuming a star formation efficiency  $\epsilon_{\rm SF}$ the 
total mass of the seed disc can be calculated. Since 
the radial extent of the discs is observationally constrained, this mass 
can be converted into a surface density $\Sigma$ of the 
proposed gas disc. The observations of the stellar discs suggest that the mass 
density decreases nearly as $1/r^{1.4}$ \citep{bartko010}, 
so we write for the mass $M_{\rm d}$  of the 
gaseous disc
\begin{equation}
M_{\rm d}  = 5360 \mu_{5360}\epsilon_{\rm SF}^{-1} M_{\odot},
\end{equation}
where $\mu_{5360}=(M_{\rm s}/5360 M_{\odot})$
and for the surface density $\Sigma_{\rm d}(r)  =  \Sigma_{0.1} \hat{r}^{-1.4}$ we find
\begin{equation}
\Sigma_{0.1}  =  9.52\times10^4\mu_{5360}\epssf^{-1}  \frac{\rm \msun}{\rm pc^2}
              = 19.89 \mu_{5360}\epssf^{-1} \frac{\rm g}{\rm cm^2},
\end{equation}
with $\hat{r}\equiv r/ 0.1{\rm pc}$. In Fig. \ref{fig:sigmaimf} 
we show for two star formation efficiencies 
the  surface density of the disc obtained as described above. The solid 
line assumes $\epsilon_{\rm SF}=0.01$, and the dotted line assumes 
$\epsilon_{\rm SF}=1$.
\subsection[Radiation Driven Warping]{Radiation Driven Warping 
and the Galactic Centre Disc}
\label{sec:radwarp}
Radiation warping of accretion discs  is studied in detail by many authors 
\citep{pringle96,maloney96,pringle97,ogilvie01}. When an optically thick disc 
is exposed to central radiation, and it re-emits the absorbed incident photons 
parallel to the disc local normal, an inward directed  force is experienced by 
each side of the disc. If the disc is slightly distorted, a net torque is induced 
which results in a modification of the warp, and precession of the disc around the 
total angular momentum direction. Whether the disc will acquire a pronounced 
warp or not depends on the competition between the net torque, and 
the component of the stress in the $r-z$ direction in cylindrical symmetry, 
present in warped discs. The latter forces the disc to settle onto a plane on 
an alignment 
time-scale. The alignment time scale at distance $r$ from the 
black hole is associated with the $r-z$ stress, and can be written as 
$t_{\nu_2} = 2 r^2/ \nu_2$, where $\nu_2$ is the vertical viscosity coefficient. 
The condition that the warp growth time scale, $t_{\rm \Gamma}$, be smaller than the 
viscous time scale, $t_{\nu_2}$ is written as \citep{pringle97}
\begin{equation}
\frac{12 \pi \Sigma r^3 \Omega c}{L} \leq  \frac{2 r^2}{\nu_2}, 
\label{eq:radpres}
\end{equation}
where $L$ is the luminosity of the central source, $c$ is the speed of light, and $\alpha$ 
is the 
Shakura-Sunyaev parameter \citep{shakura73}.  Equation (\ref{eq:radpres}) assumes 
an accretion disc which is in steady state. For such discs, the luminosity of the 
disc $L$ is related to the radiative efficiency $\epsilon \equiv L / \dot{M} c^2$, 
where $\dot{M}$ is the mass accretion given by $\dot{M} = 3 \pi \Sigma \nu_1$, and $\nu_1$ 
is the radial viscosity coefficient. In order to evaluate the warping criterion 
given in (\ref{eq:radpres}) 
one has to estimate the magnitude of the vertical viscosity $\nu_2$. 
Previous studies of linear hydrodynamic warps and magnetized shearing box simulations of
accretion discs showed that $\eta_{\nu}=\nu_2 / \nu_1 = 1 / 2 \alpha^2$  
\citep{papa83,ogilvie99, torkelsson00}. \cite{lodato07} find with smoothed particle 
hydrodynamics simulations
of small and large-amplitude warps that the vertical viscosity saturates for
small $\alpha$ such that $\eta_{\nu} < 3.5 / \alpha$. Using these, the warp 
damping time scale becomes
\begin{equation}
t_{\nu_2} = \frac{2 r^2}{\nu_2}
            \simeq \left(\frac{\alpha}{3.5}\right)
\frac{2 r^2}{\nu_1}
            \approx 5.6 \times10^5 
      {\frac{\alpha\mu_{5360}\epsilon \hat{r}^{0.6}} {\epsilon_{\rm SF}\eta_{\rm edd}}} {\rm yr},
\end{equation}
where $\eta_{\rm edd} \equiv L/L_{\rm edd}$. Equation (\ref{eq:radpres}) 
can now be rearranged to give the radiation warping critical radius
\begin{equation}
  R_{\rm rad} > {\frac{2 \eta_\nu^2}{\gamma_{\rm crit}^2 \epsilon^2}} \frac{2GM_{\rm bh}}{c^2},
\end{equation}
where $\gamma_{\rm crit}\simeq 0.32$ \citep{pringle97}.
One way to obtain a stronger torque is if the irradiation of the disc
is driving an outflow \citep{schandl94}. In this case, the
torque is determined by the outflow momentum or pressure at the sonic
point, which is basically $\propto L$, but also depends on the
detailed disc structure in a complicated way. The enhancement of the
torque could be significant, as the ratio of momentum to energy for
particles may be much larger than for photons. Although a gross over
simplification, we may parameterize this by an additional (not
constant) multiplicative factor $F_{\rm wind}$ on the radiation
torque, so that the warping criterion becomes
\be
  R_{\rm rad} > {2 \eta_\nu^2 \over \gamma_{\rm crit}^2 \epsilon^2 F_{\rm wind}^2} 
          \frac{2GM_{\rm bh}}{c^2}
       = 9.1\times10^{-3} \alpha^{-2}\epsilon_{0.1}^{-2} 
           F_{\rm wind}^{-2}\ {\rm pc}, 
\ee
where for the numerical value we have used $\gamma_{\rm crit}\simeq 0.32$,
$\epsilon=0.1\epsilon_{0.1}$ and $\eta_\nu\simeq 3.5/\alpha$. 
Thus warping the disc in the radial range of the GC discs (0.03-0.5pc) 
requires $\alpha\sim 0.3$, or $\alpha\sim 0.1$ and radiatively efficient 
accretion by a rotating black hole or modest enhancement of the torque by 
a disc wind.

\subsubsection[Warping, Irradiation and Star Formation]
{The Effect of Warping on Irradiation and Fragmentation}
\label{sec:irradiation}
Having argued for the possibility of radiation warping of the 
past accretion disc at the GC, we now discuss the feedback effect of 
warping on the irradiation and fragmentation of the disc. 

\cite{pringle97} studied the growth and non-linear evolution of 
radiation induced warps in the context of AGN accretion discs. His simulations 
show that highly warped discs exhibit limit cycle oscillations as a 
consequence of self-shadowing. As the warp gradually grows, the solid angle 
subtended by the central source decreases. When the maximum inclination of 
$\pi$ is reached, the instability cuts off leading to a decrease of the disc 
inclination. Thereafter, the disc geometry again lets the instability set 
in, and the warp grows. Hence, even though the disc might be warped by 
large angles, radiation instability keeps acting on the 
disc so long as there is sufficient amount of radiation from the central parts.  

The effect of warping on fragmentation is studied in \cite{goodman03} who 
estimated for an irradiated disc the stabilizing temperature, $T_{\rm st}$, above 
which fragmentation can not occur. When $T_{\rm st}$ is 
much larger than the temperature of the disc in equilibrium with 
radiation from the central source $T_{\rm eq}$, irradiation can not prevent 
fragmentation even for highly warped discs. The ratio $T_{\rm st} / T_{\rm eq}$ 
depends (weakly) on the mass of the central object and for a black hole of 
mass $M_{\rm bh} =  4 \times 10^6 M_{\odot}$, and a disc with a warp 
of $60^{\circ}$ it 
becomes $\sim$ 1 at 0.03 pc and $\sim$ 4 at 0.5 pc for $\alpha = 0.3$, and 
$\epsilon = 0.1$. We note that for a disc around a black hole of mass 
$M_{\rm bh} = 10^8 M_{\odot}$, this ratio would have a value  $ > 10$ at all radii, 
therefore fragmentation for a surrounding disc would be most likely. For the case of 
the GC on the other hand, the ratio  $T_{\rm st} / T_{\rm eq}$ being close to unity at the 
inner edge of the disc suggests that fragmentation might be somewhat delayed 
until the effect of irradiation on the disc diminishes 
(see  \S \ref{sec:interplay}).

\subsection[Bardeen-Petterson Effect]{Bardeen Petterson Effect and the Galactic 
Centre Stellar Disc}
\label{sec:bardeen}
In this section, we show under which conditions the Galactic Centre disc 
might have been warped due to Bardeen-Petterson effect.

An accretion disc forming around a rotating (Kerr) black hole might initially have a 
total angular momentum misaligned with that of the black hole. Inner portions of the 
disc close to the 
black hole experience general relativistic frame dragging which causes a differential 
precession, the so-called Lense-Thirring precession \citep{lense1918}. 
As we have seen in the previous section, the viscous time scale increases with
radius, hence the inner parts of the disc are forced  to align with the black hole 
on time scales much shorter than those for the outer parts. Consequently, the disc 
develops a shape which for $r < R_{\rm BP}$ is aligned with the black hole, i.e. is 
flat, and for $r \geq R_{\rm BP}$ its angular momentum direction changes gradually 
from radius to radius \citep{bardeen75}. Like the radiation pressure warping, 
the Bardeen-Petterson effect also is a competition between the alignment time scale, 
and the precession time scale \citep{armitage99}. The precession induced by the 
Bardeen-Petterson effect 
is given by \citep{kumar85}
\begin{equation}
\dot{\phi}=2 a c \left( \frac{G M_{\rm bh}}{c^2}\right)^2 \frac{1}{r^3},
\end{equation}
where $a$ is the black hole spin parameter which can take values between $0$ and $1$ 
for stationary, and maximally rotating black holes respectively. The precession 
time scale $\tau_{\rm BP} = 2 \pi / \dot{\phi}$ is
\begin{equation}
\tau_{\rm BP} = \frac{\pi c^3 r^3}{a G^2 M_{\rm bh}^2}.
\label{eq:tbard}
\end{equation}
The critical radius, where the alignment time scale equals the precession time
scale is then obtained by writing
\begin{equation}
\frac{\pi r^3 c^3}{a G^2 M_{\rm bh}^2} = \frac{2 r^2}{\nu_2} 
\rightarrow R_{\rm BP}= \frac{2 a G^2 M_{\rm bh}^2}{\pi c^3 \nu_2}.
\label{eq:bardeen}
\end{equation}
Using the steady state relations introduced in the previous section equation 
(\ref{eq:bardeen}) becomes
\begin{equation}
R_{\rm BP}=\frac{6 a G^2 M_{\rm bh}^2 \epsilon \Sigma(R_{\rm BP})}{\eta_{\nu} c L}.
\label{eq:bcrit}
\end{equation}
In order to estimate the Bardeen-Petterson radius for the 
GC, we need to solve equation (\ref{eq:bcrit}) for $R_{\rm BP}$. 
Writing $\Sigma(R_{\rm BP})=\Sigma_0/ \hat{r}^{-1.4}$ we obtain
\be
R_{\rm BP}=4.22 \times 10^{-4} \left( \frac{a \epsilon \alpha 
\mu_{\rm 5360}}{\eta_{\rm edd} \epsilon_{\rm SF}} \right)^{1/2.4} \ \rm{pc}.
\ee
This estimate shows that the Bardeen-Petterson radius for the GC 
is quite small, i.e. much below the inner edge of the 
observed discs. We should remind that the alignment 
time scale $t_{\nu_2}$ for the assumed surface density profile is short.
This means that the disc can be warped out to large distances 
since both $R_{\rm BP}$ and $t_{\nu_2}$ depend on the disc 
parameters only weakly. 

The angular momentum of the disc at $r$ is $\propto 2\pi\Sigma rdr 
\sqrt{GM_{\rm bh}r} \propto r^{0.1} dr$ and varies only slowly
with $r$. The ratio of the disc to black hole angular momenta can
thus be estimated as
\be
 {J_d\over J_{\rm bh}} \simeq {M_d \sqrt{GM_{\rm bh}r_d}
      \over aGM_{\rm bh}^2/c} = \frac{1}{a} \frac{M_d}{M_{\rm bh}}
      \sqrt{\frac{2r_d}{r_s}} \sim 0.97 \epsilon_{\rm SF}^{-1} a^{-1},  
\ee
where $r_{\rm s}=2 G M_{\rm bh}/c^2$ is the Schwarzschild radius, and 
$r_{\rm d}\sim 0.1$ pc. This suggests that by the time the 
disc is significantly warped at
$r_{\rm d}\sim0.1$ pc, the black hole spin should be significantly
aligned: the alignment time-scale found by \cite{lodato06}
under similar circumstances is $\sim 0.3 t_{\nu_1}\sim 0.3
(3.5/\alpha) t_{\nu_2} \sim \alpha^{-1} t_{\nu_2}$.  These
arguments suggest that if the GC disc was first warped through the
Bardeen-Petterson effect, and then fragmented to form the observed
surface density of young stars, the star formation efficiency had to
be high, $\epsilon_{\rm SF} <  1$, to prevent the disc from
dominating the angular momentum, and $\alpha<0.1$ to give the
disc time to warp before it is accreted and will align the black
hole spin.

\subsection[Comparison of Gravitational, Viscous and Radiation Torques]
{Comparison of Gravitational, Viscous  and Radiation Torques}
\label{sec:comparetrq}

In \S \ref{sec:radwarp} we have discussed the conditions for radiation warping. 
While equation (\ref{eq:radpres}) has to be satisfied for warping, 
it is also useful to compare the magnitude of radiation and 
gravitational torques acting on the disc.

The magnitude of the radiation torque on a ring of radius $r$ with
radial width $dr$ can be approximated as \citep{ogilvie01}
\be
2\pi r dr T_\Gamma  \simeq {\frac{L}{6 c r}} r dr, 
\ee
thus $T_\Gamma=\Sigma r^2 \Omega/t_\Gamma$. 
On the other hand the 
viscous torque on the ring, which tries to damp the warp in the case
of radiation pressure warping, and which sets up the warped density 
distribution in
the Bardeen-Petterson mechanism, is given by
\begin{equation}
2\pi r dr T_{\nu_2}  =  {\Sigma r^2\Omega 2\pi r dr \over 2r^2/\nu_2 } 
                    =  {\eta_\nu L \Omega r dr \over 3 \epsilon c^2}.
\end{equation}
The gravitational torque of the disc on the same ring can
be determined by integrating over the disc
\begin{equation}
2\pi r dr T_{\rm grav} 
   \simeq {G M_{d} \Sigma(r) 2\pi r dr \over r} J, 
\end{equation}
where $J \equiv T_{\rm grav} / (G m_i M_d / r_i )$ is a dimensionless integral 
depending on $r/r_1$, $r/r_2$
\begin{equation}
 J =  \int_{r_1}^{r_2} {2\pi r' dr' \Sigma(r') \over M_d} 
        {r^2 r'\over (r'^2+r^2)^{3/2}} \sin(2\beta) \, I(\beta,r/r')
        {\partial\beta\over\partial\theta},    
\end{equation}
and where $\beta(r,r')$ is the angle between the normals of the two
rings at $r$ and $r'$, $\theta$ is the inclination of the ring at $r$, 
so $\partial\beta/\partial\theta=O(1)$,
and $I(\beta,r/r')$ is the integral in equation (13a) of 
\cite{ulubay-siddiki09} (hereafter US09). 

\begin{figure}
\centering
\includegraphics[width=8.5cm]{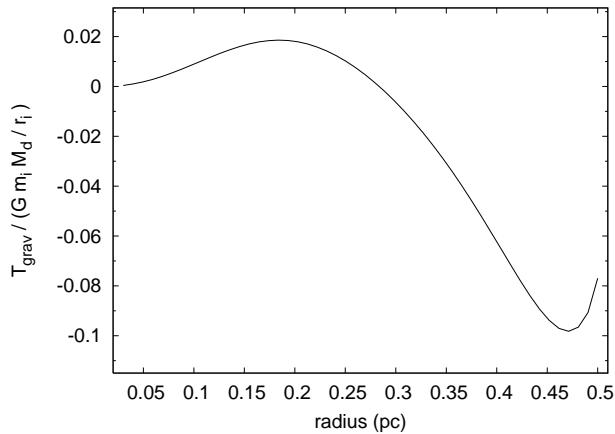}
\caption[Normalized gravitational torque for the GC disc]
{Normalized gravitational torque for the GC disc when 
the warp spans a range of $-15^{\deg}\rightarrow 15^{\deg}$ in inclination.}
\label{fig:Jterm}
\end{figure}

We show in Fig. \ref{fig:Jterm} 
the values of the $J$-terms at different 
distances from the black hole when the warp spans a range of 
$-15^{\deg}\rightarrow 15^{\deg}$ in inclination. For larger warps, the 
gravitational torques are weaker. Also if we include the part of the disc 
outside $[r_{\rm in}, r_{\rm out}]$ the product $M_{\rm d}J$ decreases as the torques 
are dominated by the nearby parts of the disc. 
Using the surface density profile above, the ratio of the radiation
and gravity torques is 
\be
{T_\Gamma \over T_{\rm grav}} =  {L\over 6 c r} 
            {r \over  2\pi GM_{\rm d}\Sigma(r) J} 
    = 32.5 {\eta_{\rm edd} \epssf^2 \over \mu_{5360}^2} {\hat{r}^{1.4}\over J}, 
\label{eq:radgrav}
\ee
and the ratio of the viscous torque to the gravitational torque is
\be
{T_{\nu_2} \over T_{\rm grav}} = {\eta_\nu L \Omega \over 3 \epsilon c^2}
            {r \over  2\pi GM_{\rm d}\Sigma(r) J} 
    = 3.14  {\eta_{\rm edd} \epssf^2 \over \mu_{5360}^2\alpha\epsilon_{\rm 0.1}} 
               {\hat{r}^{0.9}\over J}. 
\label{eq:visgrav}
\ee
We see from Fig. \ref{fig:Jterm} that the $J$-terms for the Galactic 
Centre disc surface density profile are 
typically $O$(0.1). Thus at $r=0.1$ pc $T_{\Gamma}\gg T_{\rm grav}$ if 
$\epsilon_{\rm SF} < 1$, but not if $\epsilon_{\rm SF}\ll 1$. 
For a high central luminosity $\eta_{\rm edd}$ 
and/or high star formation efficiency $\epsilon_{\rm SF}$ the viscous torque on the 
disc dominates over its self-gravity. We have also seen for the case of 
the Bardeen-Petterson effect that $\epsilon_{\rm SF} < 1$. This 
shows that it is justified to 
neglect the effects of the gravitational torques on the evolution of the 
disc in its active phase.
 
\subsection{The Warped Stellar Disc After Fragmentation}
\label{sec:interplay}
In sections (\ref{sec:radwarp}) and (\ref{sec:bardeen}) we have seen that
for a range of assumed accretion disc parameters, the disc 
at the GC could have been warped in the accretion phase. 
The evolution of the disc is likely to be governed 
by the viscous or radiation torques during this phase. 

We now assume that some time after the disc has become warped, 
the active phase of the GC ends. Since $\eta_{\rm edd}$ and the 
accretion rate will then be highly reduced the disc will 
receive much smaller energy input and it is reasonable to assume that 
it can now cool rapidly. If the cooling time is short,  
fragmentation and star formation can occur on a 
dynamical time scale \citep{gammie01}. The gas content of the disc could 
then  be converted into stars with efficiency $\epsilon_{\rm SF}$ and the 
rest of the gas will be lost.

Subsequently, the now stellar disc would be subject only
to gravitational forces, and precess 
accordingly. In addition to the precession caused by the disc's self-gravity, the 
surrounding old star cluster will also contribute unless it is 
spherically symmetric. After analyzing the proper motions of the old cluster 
stars, \cite{trippe08} reported a small amount of rotation 
in the direction parallel to the Galactic rotation. The rotation 
of the old cluster is also confirmed by \cite{schoedel09}. Based on 
these findings, one might consider the possibility of the star 
cluster being slightly flattened, hence have an effect on the long term 
evolution of the disc.

\section{Warp Model for the Galactic Centre Disc}
\label{sec:model}
In the previous sections  we have argued that for a range of parameters,  
a flat disc  around SgrA* could become warped before forming stars, 
after which the disc evolves gravitationally. In this section, 
we will introduce the model we have used to follow the 
time evolution of the warped stellar disc. 

\subsection{The Equations of Motion}
\label{sec:eqnsofmot}

As in US09, we model a warped disc as a collection of concentric 
circular rings which are tilted with respect to each other. 
The rings are in gravitational interaction with each other, and are also 
torqued by the surrounding old star cluster.  
They are characterized by 
their masses $m_i$, and radii $r_i$. The fast orbital motion around 
the SMBH of mass $M_{\rm bh}$ is thereby time-averaged. The 
geometry of the rings is defined by the Euler angles $(\psi, \theta, \phi)$.
For a system of $n$ rings, the equations of motion for any 
of the rings $i$ are given by
\begin{equation}
p_{\theta_i}=\frac{m_i r_i^{2}}{2}\dot{\theta_i},
\label{eq:three}
\end{equation}
\begin{equation}
p_{\phi_i}= \frac{m_i r_i^{2}}{2}\dot{\phi_i} \sin^{2}\theta_i +p_{\psi_i} \cos\theta_i,
\label{eq:four}
\end{equation}
\begin{equation}
 \dot{p_{\theta_i}}=\frac{m_i r_i^{2}}{2} \dot{\phi}_i^{2} \sin \theta_i \cos \theta_i 
-\dot{\phi}_i p_{\psi_i}
\sin \theta_i - \frac{\partial V_i}{\partial \theta_i},
    \label{eq:two}
 \end{equation}
\begin{equation}
\dot{p_{\phi}}_i=-\frac{\partial V_i}{\partial \phi_i}.
\label{eq:six}
\end{equation}
Here $p_{\psi_i}=m_i r_i^2 \Omega_i$ is the orbital angular momentum,  
$\dot{\phi_i}$ and $\dot{\theta_i}$ are the rates of 
precession and nutation caused by the torques 
$\partial V_i/ \partial \theta_i$ and $\partial V_i/ \partial \phi_i$
respectively, where $V_i$ is the potential energy of ring $i$ in the 
field of the other rings, and of the old stellar cluster. The details 
of the calculation of the mutual torques can be found in US09. To calculate 
the torque due to the surrounding star cluster, we use the description 
given in \cite{sparke86} appropriate for slightly flattened systems. 
The potential energy of a ring in the field of the cluster 
is given by 
\begin{equation}
V_{\rm cl}=\frac{m_i v_0^2}{6} (1-q) \sin^2 \theta_i,
\end{equation}
where $v_0$ is the assumed constant circular velocity generated by 
the star cluster, and $q$ is the flattening. The torque on ring $i$ 
due to the star cluster is then written as
\begin{equation}
\frac{\partial V_{\rm cl}}{\partial \theta_i}=\frac{m_i v_0^2}{6} (1-q) \sin 2 \theta_i. 
\label{eq:clstrq}
\end{equation}
The precession frequency of the disc induced by the star cluster 
at different radii $r$ is given by
\begin{equation}
\dot{\phi}_i=-\frac{v_0}{3 r_i}(1-q) \cos \theta_i.
\label{eq:clusterphdt}
\end{equation}
\subsection{Numerical Setup}
\label{sec:setup}
\begin{figure}
\begin{centering}
\includegraphics[width=8.5cm]{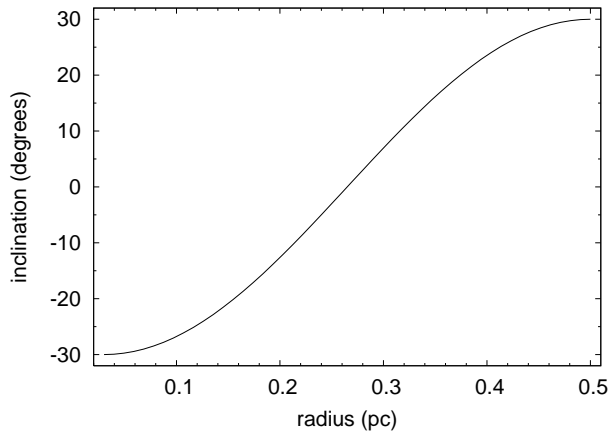}
\caption{Initial inclination profile of the model discs.}
\label{fig:theta_init}
\end{centering}
\end{figure}

The numerical approach we use to study the time evolution of 
the disc is similar to the one presented in US09. 
There, we integrated the equations of motion for the rings 
after evaluating the steadily precessing equilibria for the given 
parameters to test the stability of those configurations. 
For the parameters relevant to the GC disc we do not find such steadily 
precessing equilibria. In an earlier work it is shown that 
centrally illuminated discs around black holes might under certain 
circumstances be warped by large angles, e.g. even by 
$\theta \sim \pi$ \citep{pringle97}. The final 
geometry of the warped disc depends on the model assumptions like the strength of the 
perturbation leading to radiation instability and the location in the disc where the 
perturbation occurs. For the case of Bardeen-Petterson warping on the other hand, 
the degree of the warp depends on the initial mutual inclination between the 
angular momenta of the gaseous disc and the black hole. 
Therefore the most realistic initial disc shape for our simulations 
can only be chosen after carrying out simulations of gaseous discs subject 
to radiation instability, or Bardeen-Petterson effect, which is beyond the 
scope of this study. Therefore, without attempting to find such 
precise initial conditions, we adopt as initial warp shapes, the 
disc configurations which are likely to be imposed 
by the warping mechanisms discussed in the previous sections. Specifically, all 
the models are assigned an initial warp profile such that the inclinations 
range from $-30^{\circ}$ to $30^{\circ}$, and the azimuthal angles are zero. 
We note that this choice of the inclinations is in the range 
of the so-far reported values for either of the mechanisms. 
Fig. \ref{fig:theta_init} depicts the initial inclination 
profile for the models we constructed. 

\begin{table*}
\begin{center}
\begin{tabular}{l c c c c c c c}
\hline
Simulation & $M_{\rm d} /M_{\rm bh} $ & $r_{\rm in}-r_{\rm out}$ & $\Delta \theta$ &  $|\dot{\phi} / \Omega_{\rm in }|$ & $  v_0^2 (1-q) $     & $\rm \tau_{p}$    & Disc  \\
           &                      & (pc)                   & (degrees)     &                                & $\rm (km^2 \ s^{-2})$ & $\rm (10^6 \ yr)$ & Break Up \\
\hline
  $\rm SG\_HM$   &   0.13400      & {\underline{0.03-0.5}}    & {\underline{60}}    & $10^{-5}$ & -   & 0.6  & yes \\
  $\rm SG\_IM$   &   0.01340      & {\underline{0.03-0.5}}    & {\underline{60}}    & $10^{-5}$ & -   & 8    & yes \\
  $\rm SG\_LM$   &   {\underline{0.00134}}  & {\underline{0.03-0.5}} & {\underline{60}} & $10^{-5}$ & -   & 60   & no  \\
  $\rm CLSG\_HM$ &   0.13400      & {\underline{0.03-0.5}}    & {\underline{60}}    & $10^{-5}$ & {\underline{860}} & 0.35 & yes \\
  $\rm CLSG\_IM$ &   0.01340      & {\underline{0.03-0.5}}    & {\underline{60}}    & $10^{-5}$ & {\underline{860}} & 0.8  & yes \\
  $\rm CLSG\_LM$ &   {\underline{0.00134}}  & {\underline{0.03-0.5}} & {\underline{60}} & $10^{-5}$ & {\underline{860}} & 1.6  & no  \\
\hline
\end{tabular}
\caption{Parameters of the simulations discussed in the paper. The first 
column shows the label of the simulation. SG models are those considering only 
the self-gravity of the disc, while CLSG models include the effects of the star 
cluster as well. HM, IM, and LM stand for high mass, intermediate mass, and low 
mass respectively, which refer to the mass fractions given in column two. The third, 
the fourth, and the fifth columns show the adopted radial range, the 
overall warp, and the normalized initial precession frequency respectively. The sixth column 
gives the value of the quantity $v_0^2 (1-q)$ which measures the strength of 
the torque from the star cluster. The seventh column lists the precession times attained 
at the end of each simulation, where the last column summarizes the behavior of the 
disc in terms of its coherence. The underlined parameter values follow from 
the observations or the models, whereas the others are assumed values.}
\label{table:params}
\end{center}
\end{table*}

In our simulations, we model a disc as a collection of 50 equally spaced, circular 
rings. The model discs extend between $0.03-0.5$ pc, surrounding a black hole of 
mass $M_{\rm bh}=4 \times 10^{6} M_{\odot}$. The surface density of the discs 
decline as $1/r^{1.4}$ as indicated by the observations. Initially, the discs are 
given a precession frequency such that when normalized to the orbital 
frequency of the innermost ring it gives $\rm \dot{\phi} / \Omega_{in}=-10^{-5}$. 
Below we will often give evolution times in terms of the
orbital time at the inner edge of the disc, $\rm (P_{orb})_{in} \approx 244$ yr.
We consider discs which are under the influence of only 
self-gravity (models SG), and discs which experience 
the effects of the star cluster and the self-gravity 
together (models CLSG). We assume three different mass 
fractions, $  M_{\rm d} /M_{\rm bh} = $0.00134, 0.0134, and 
0.134 (LM-low mass, IM-intermediate mass, and HM-high mass respectively). 
We note that assuming $M_{\rm d} /M_{\rm bh}= 0.00134$ implies 
a star formation efficiency of $\epsilon_{\rm SF} = 1$, a value for which 
warping the planar GC disc is most plausible (see \S \ref{sec:warpingGC}). 
However, in order to keep our discussion broad, and to make our models 
applicable to systems other than the GC, we also explore mass fractions 
which are higher than that inferred from the observations of the GC discs.

Observations of the nuclear star cluster of the GC suggest a mass 
range of $M_{\star} \sim [1-2]\times 10^6 M_{\odot}$ within the 
outer edge of the discs at 0.5 pc  \citep{trippe08, schoedel09}. Therefore, 
for the CLSG models, we set the quantity $v_0^2(1-q)$ that determines 
the strength of the cluster torque to 
860 $\rm km^2 \ s^{-2}$ (see equation (\ref{eq:clstrq})). 
This value can be obtained, for example by setting 
the cluster mass within 0.5 pc to $10^6 M_{\odot}$, 
hence adopting a circular velocity of $v_0\sim 93$ km/s as induced 
by the star cluster, and the flattening of the cluster to $q=0.9$. 

The parameters of the simulations along with the resulting precession 
time scales and evolution are listed in Table \ref{table:params}. 
In these simulations, the total energy of the rings is typically 
conserved to an accuracy of $10^{-5}$.

\section{Results}
\label{sec:results}
In this Section, we describe the main results that emerged from 
our simulations. We discuss in turn the precession of the model discs and 
the evolution of the discs' warp shapes. 
\subsection{Precession of the Disc}
The self-gravity and the star cluster torques will force the 
disc to precess on time-scales proportional 
to the disc mass fraction, and the quantity $v_0^2(1-q)$ respectively 
(see equation (28) in US09 for the case of purely self-gravitating discs). The 
overall evolution of the disc will then be 
determined by the relative strengths of both torques.

\begin{figure}
\begin{centering}
\includegraphics[width=8.2cm, height=6cm]{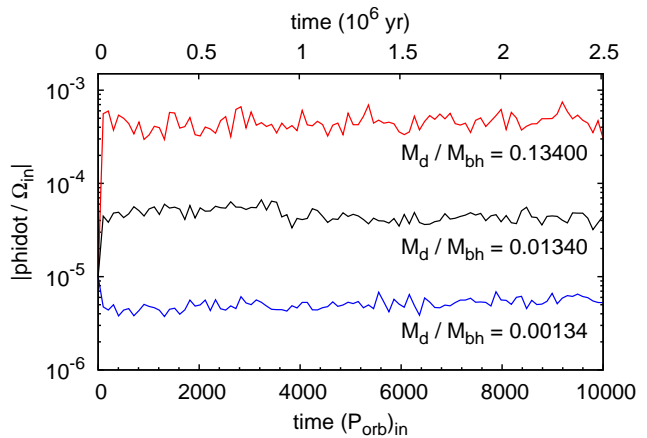}
\caption{Evolution of the precession frequencies of the models 
considering only self-gravity of the disc. Different curves correspond to 
models with different mass fractions. The precession frequency 
increases in proportion to the disc mass fraction.}
\label{fig:SG_phidots}
\end{centering}
\end{figure}

In Fig. \ref{fig:SG_phidots} we show for the SG models how the 
precession frequencies evolve during the simulations. 
The x-axis shows the elapsed time in terms of the orbital frequency of 
the innermost ring on the lower x-axis, and in units of $10^6$ yr on the 
upper x-axis. The y-axis shows the absolute value of the averaged 
precession frequency, i.e. 
$|\displaystyle\sum_{i=1}^{n} \dot{\phi}_i / n |$. The red (upper), 
the black (middle) and the blue (lower) curves correspond to disc mass 
fractions of $ M_{\rm d} / M_{\rm bh} = 0.13400$, 
$M_{\rm d} / M_{\rm bh}  = 0.01340$, and $M_{\rm d} / M_{\rm bh}= 0.00134$ 
respectively. We see that although 
the average precession frequencies exhibit small oscillations in amplitude, 
they attain their true values (imposed by the disc's gravity) 
soon after the simulation starts, remaining approximately constant 
even though the shape of the disc evolves, as is shown in the next 
subsection. While individual rings may precess in the forward direction
for short time intervals, most rings precess backwards for
most of the time, so that the average precession is always
retrograde.

Throughout the simulation model $\rm SG\_HM$ precesses with 
a normalized frequency of about $-4 \times 10^{-4}$, corresponding to a 
precession time scale of $\rm \tau_{p} \approx 6 \times 10^5 $ yr. 
For model $\rm SG\_IM$ the normalized precession frequency
drops to approximately $-3 \times 10^{-5}$ giving 
$\rm \tau_{p} \approx 8 \times 10^6 $ yr, 
and when the mass fraction is decreased even further we see from model 
$\rm SG\_LM$ that $\rm \dot{\phi} / \Omega_{in} = -4 \times 10^{-6}$ leading 
to $\rm \tau_{p} \approx 6 \times 10^7 $ yr. 
(In order to check this, we have computed for model $\rm SG\_LM$ the normalized precession 
frequencies that the rings would have under the self-gravity torques when 
the initial warp profile is considered (see equation (14) in US09). The values are 
in the range $[-8,-2] \times 10^{-6}$, with some of the rings 
having $ \dot{\phi}/\Omega_{\rm in}\sim -4 \times 10^{-6}$, consistent with 
Fig. \ref{fig:SG_phidots}.) We can readily conclude 
from this, that the precession frequencies scale with disc mass. 
In the equations of motion,  when the quadratic term for the 
precession frequency is set to zero, all terms scale 
linearly with ring mass so the precession frequency should indeed be proportional 
to the mass. Among the models discussed above only $\rm SG\_HM$ would have completed a 
full precession during the life time of the stars. 

\begin{figure}
\begin{centering}
\includegraphics[width=8.2cm, height=6cm]{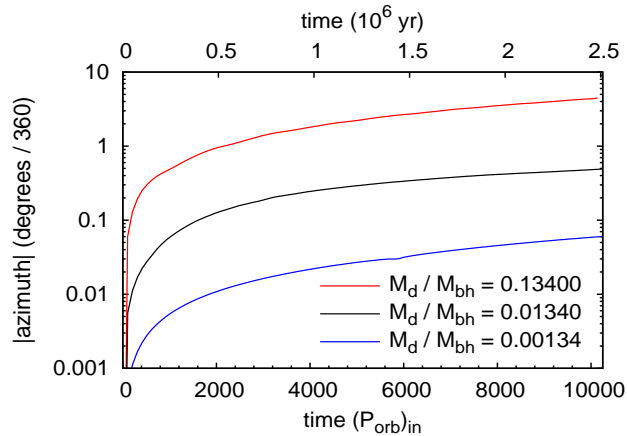}
\caption{Precession of the models considering only self-gravity of the disc. 
Different curves correspond to the models with different mass fractions: 
the red (upper) curve  is for model $\rm SG\_HM$, the black (middle) 
curve  is for model $\rm SG\_IM$, 
and the blue (lower) curve  is for model $\rm SG\_LM$. The ever faster precession of the 
more massive discs can be clearly seen.} 
\label{fig:SG_phis}
\end{centering}
\end{figure}

In  Fig. \ref{fig:SG_phis} we display how the above scaling manifests 
itself when we look at the averaged azimuthal angles of the rings, 
where the averaging is done in a similar way as for the frequencies. 
The different curves correspond to the different mass fractions again. 
At the end of the simulation, 10000 orbital periods after 
the simulation has started, model $\rm SG\_HM$ is seen to have precessed 
approximately 4.5 times, while model $\rm SG\_IM$ has completed nearly 
half a precession. The precession of the least massive model $\rm SG\_LM$
on the other hand can barely be noticed with 0.05 times a full precession. 

\begin{figure}
\begin{centering}
\includegraphics[width=8.25cm, height=6cm]{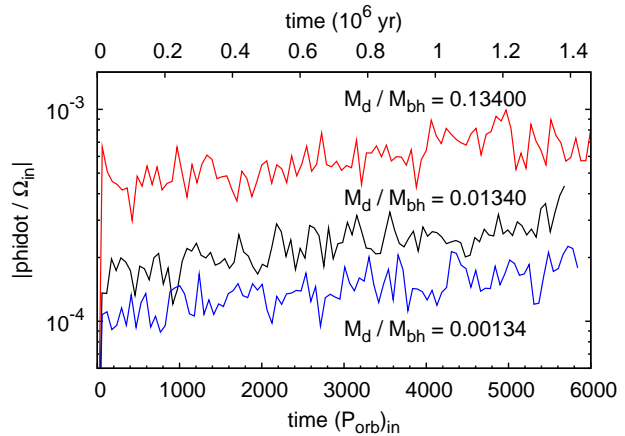}
\caption{Evolution of the precession frequencies of the models 
including the torques from self-gravity and from the surrounding star cluster. Different curves 
correspond to models with different mass fractions as indicated on the labels.} 
\label{fig:CLSG_phidots}
\end{centering}
\end{figure}

We now move on to the models including the effect of a surrounding 
cluster. In Fig. \ref{fig:CLSG_phidots} we show the evolution of 
the averaged precession frequencies for the CLSG models. In contrast to 
the SG models, the precession frequencies for the CLSG models may not 
reach a constant value (note that the noise in the curves is the same as 
in Fig. \ref{fig:SG_phidots}, given the axis scaling). Until 3000 orbital periods, 
the frequencies of all the models increase linearly in time. Since the 
parameters of the star cluster are the same for all the 
models, it is again the most massive disc which precesses with 
the highest rate $\rm \dot{\phi} / \Omega_{in} = -7 \times 10^{-4}$  
corresponding to a precession time scale $\tau_p \approx 3.5 \times 10^5$ yr. 
Comparing models $\rm SG\_HM$  and $\rm CLSG\_HM$, we see that 
the averaged precession rates are approximately 
the same for the first 1000 orbital periods. 
At the end of the simulations, the star cluster 
leads to an approximately 2 times faster precession 
than in the self-gravitating case when 
$M_{\rm d} / M_{\rm bh}=0.134$. For model $\rm CLSG\_IM$ the 
precession is slower due to the decreased disc 
mass, but attains a value of 
$\rm \dot{\phi} / \Omega_{in} = -3 \times 10^{-4}$ giving 
$\tau_p \approx  8 \times 10^5$ yr, 10 times shorter 
than for model $\rm SG\_IM$ due to the influence of the star cluster. 
Finally the lowest mass disc model including the star 
cluster $\rm CLSG\_LM$ has
$\rm \dot{\phi} / \Omega_{in} = -1.5 \times 10^{-4}$ and 
$\tau_p \approx 1.6 \times 10^6$ yr, which is nearly 40 times shorter 
than for $\rm SG\_LM$. 

\begin{figure}
\begin{centering}
\includegraphics[width=8.75cm, height=6cm]{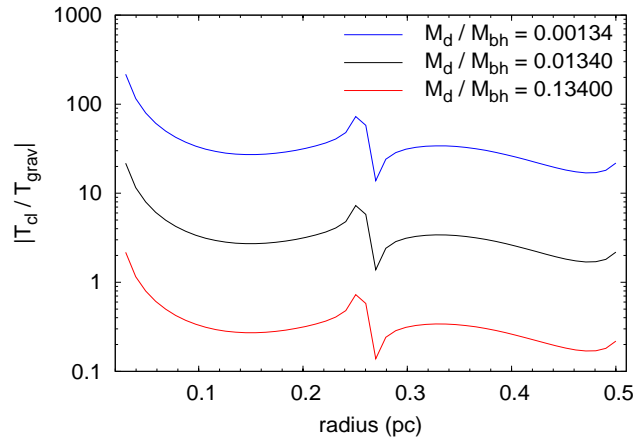}
\caption{The comparison of the cluster torques, $\rm T_{cl}$, to 
the self-gravity torques $\rm T_{grav}$ acting on the rings. The low 
mass disc is dominated by the torques from the star cluster, while 
the high mass disc is dominated by self-gravity torques.}
\label{fig:compare_trqs}
\end{centering}
\end{figure}

It emerges from these comparisons that the effect of the star 
cluster on the disc is most pronounced when the disc mass is 
low, otherwise the self-gravity torques dominate the evolution 
of the disc.
In Fig. \ref{fig:compare_trqs} we show for different 
mass fractions the expected ratio 
(based on equation (\ref{eq:clstrq}) and equations (13) in US09) 
of the cluster torque, $\rm T_{cl}$, to the self-gravity torque, 
$\rm T_{cl}$, at different locations in the disc when the initial warp 
profile is considered. One can see that the low 
mass disc is indeed dominated by the cluster, while 
the high mass disc is dominated by self-gravity except in the inner parts. 
The kink and drop around $\sim$ 0.25 pc is 
caused by the change of sign of the torques which follows from the 
choice of the warp profile.

\begin{figure}
\begin{centering}
\includegraphics[width=8.5cm, height=6cm]{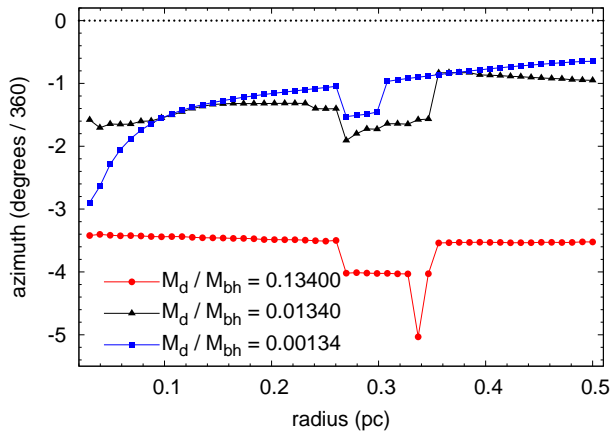}
\caption{Azimuthal profiles of the CLSG  models showing precession 
angles for all rings after 6000 inner periods.
The black horizontal dotted line at the top shows the initial configuration.
The red curve with filled circles shows the azimuthal 
profile of model $\rm CLSG\_HM$.
The black curve with filled triangles, and the blue curve with filled squares 
depict the profiles of models $\rm CLSG\_IM$, and $\rm CLSG\_LM$ respectively. 
The effect of the star cluster on the disc is best observed 
for model $\rm CLSG\_LM$ where the azimuthal angles exhibit a differential 
profile.} 
\label{fig:CLSG_radphi}
\end{centering}
\end{figure}

The relative effects of the cluster and the self-gravity torques 
can also be observed in the azimuthal profiles, i.e. radius versus 
azimuth, of the CLSG models which we 
show in Fig. \ref{fig:CLSG_radphi}. The black horizontal 
dotted line shows the initial profile of the 
azimuthal angles. The red curve with filled circles shows the azimuthal 
profile of model $\rm CLSG\_HM$ towards the end of the simulation. 
The black curve with filled triangles, and the blue curve with filled squares depict the profiles  
of models $\rm CLSG\_IM$, and $\rm CLSG\_LM$ respectively. 

In general, it is possible to find 
disc configurations where the azimuthal angles of the rings are 
aligned even in the presence of a surrounding star cluster (US09), 
however this can only be achieved for certain combinations of the 
disc parameters. Although for the GC parameters such configurations were 
not found, we see from Fig. \ref{fig:CLSG_radphi} that an almost steady 
precession can be achieved when the disc is massive enough as in model 
$\rm CLSG\_HM$. In this case, the inner and the outer parts of the disc 
are locked together, while the middle parts which are ahead of the 
others are also seen to precess together. For model $\rm CLSG\_IM$ where 
the effect of the self-gravity torques is diminished, the star cluster 
torques start to dominate in the inner parts of the disc, where the azimuthal 
profiles show more of a differential profile rather than a flat one. Since 
the rate of precession induced by the star cluster decreases going outwards 
in the disc (see equation (\ref{eq:clusterphdt})), the outer 
parts for this mass fraction are still held by the self-gravity 
of the disc. For the lowest mass model $\rm CLSG\_LM$, we see 
that the star cluster torques dominate over the self-gravity 
torques and the rings making up the disc can not precess with a 
common azimuth.

\subsection{Evolution of the Inclinations and Breaking Up of the Disc}
In this section we will discuss how the initial warp shape of the 
disc is modified under the self-gravity and the star cluster torques. Again, 
if we had started the simulations adopting equilibrium inclinations, we could 
expect to see that for a range of parameters, the discs would be stable, 
hence would not change their inclinations. Since we have seen in the previous 
section that even in the absence of an equilibrium configuration, some discs can 
maintain a near steady precession -at least at the inner and outer parts of the 
disc-, we might expect to see some behavior mimicking equilibrium in the 
the long term evolution of the inclinations as well. 

In the plots we will be depicting in this section, the x-axis will show 
the elapsed time in terms of the orbital period of the innermost ring on the 
lower x-axis, and in units of $10^6$ yr on the upper x-axis. The y-axis 
will show the ring inclinations, where the negative ones correspond to the 
inner parts, and the positive ones correspond to the outer parts. At any instant 
of time $t$, a vertical cut through the plot will then give the warp 
profile, i.e. the inclination at all radii. 

\begin{figure}
\begin{centering}
\includegraphics[width=8.5cm, height=6.cm]{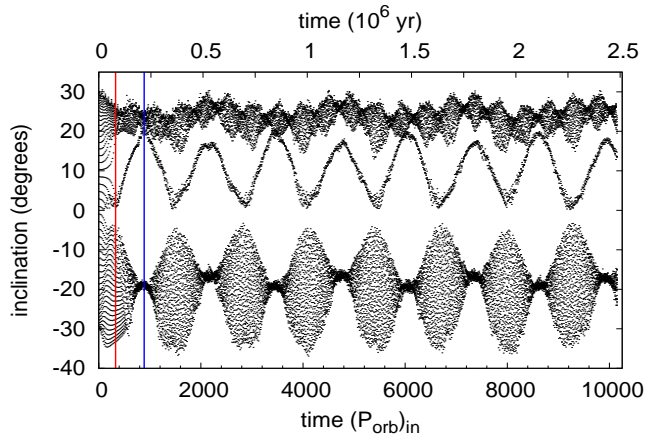}
\caption{Time evolution of the inclinations for model $\rm SG\_HM$. The 
disc breaks into separate pieces which become pronounced after 
330 orbital periods (marked by the vertical red (left) line). After 890 
orbital periods (marked by the vertical blue (right) line) the inclinations 
reach a maximum separation of about $40^{\circ}$.} 
\label{fig:SGHM_thevol}
\end{centering}
\end{figure}

We start our discussion by considering first the SG models. In Fig. 
\ref{fig:SGHM_thevol} we show the evolution of the individual ring 
inclinations for model $\rm SG\_HM$. We see that soon after the simulation 
starts the disc breaks into pieces, where the rings lying in the 
middle parts are pushed to more polar orbits, and the edge rings become 
more equatorial. After 330 orbital periods (indicated by the red (left) 
vertical line), the breaking of the disc becomes very prominent, leading to a 
configuration of three concentric, separate discs, the middle one being the 
less warped one.  
At about 890 orbital periods (indicated by the blue (right) vertical line), the 
two positively inclined discs merge together, and the separation between 
the inner and the outer discs reaches a 
maximum of about $\rm 40^{\circ}$. At this phase, the inner disc becomes 
almost planar, while the outer one is warped by about $10^{\circ}$. 
At about 1480 orbital periods the inclinations of the inner rings 
approach their initial values, and the mutual inclination of the 
separated discs become $\sim 25^{\circ}$. 
After this time, subsequent oscillations with almost equal 
amplitude and period follow. Although a more 
detailed investigation of this behavior is beyond the scope of this 
study, we should note that these oscillations are 
reminiscent of stable oscillations for which 
the system parameters oscillate around their 
equilibrium points once perturbed.

\begin{figure}
\begin{centering}
\includegraphics[width=8.5cm, height=6cm]{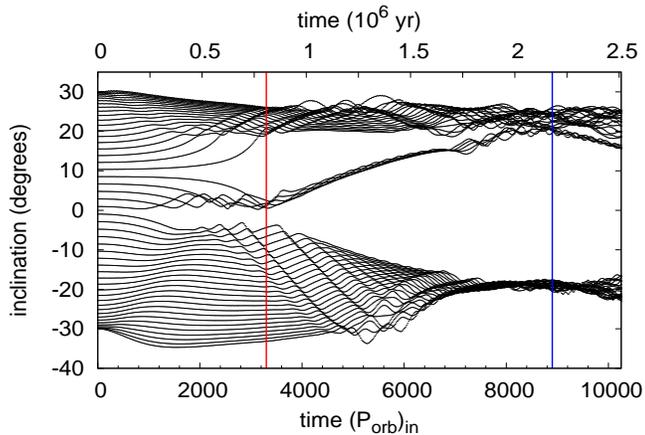}
\caption{Time evolution of the inclinations for model $\rm SG\_IM$. 
The evolution is very similar, but retarded by a factor of 10, 
with respect to the evolution of model $\rm SG\_HM$.}
\label{fig:SGIM_thevol}
\end{centering}
\end{figure}

In Fig. \ref{fig:SGIM_thevol} we show the warp evolution of model 
$\rm SG\_IM$. We see again that the disc breaks into pieces. For this model 
however the breaking becomes pronounced at about 3300 orbital periods 
(indicated by the red (left) vertical line), 10 times longer than 
for model $\rm CLSG\_HM$ corresponding to the 10 times longer precession time. 
At this stage, the disc has broken into three 
pieces, with similar characteristics, i.e. the degree of the warps, 
and the separation between the other discs, to those 
in model $\rm CLSG\_HM$. At about 8900 orbital periods 
(indicated by the blue (right) vertical line), the two of the positively inclined 
discs combine to form a single one, which  is again separated 
by $40^{\circ}$ from the negatively inclined one.  

\begin{figure}
\begin{centering}
\includegraphics[width=8.5cm, height=6cm]{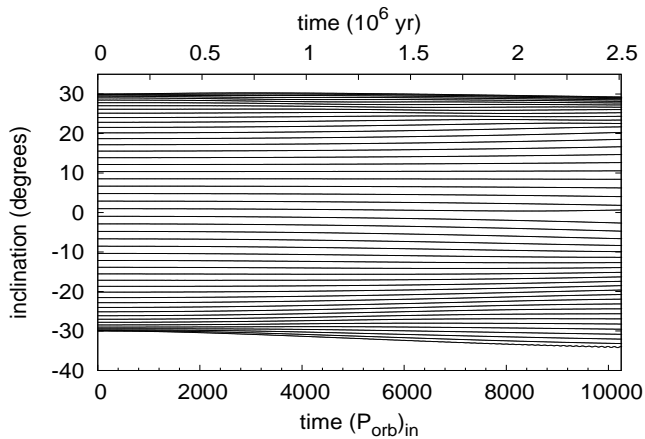}
\caption{Time evolution of the inclinations for model $\rm SG\_LM$. The 
inclinations stay almost unaltered during the simulation which 
is an effect of the low mass fraction adopted.}
\label{fig:SGLM_thevol}
\end{centering}
\end{figure}

The warp evolution of model $\rm SG\_LM$ is shown 
in Fig. \ref{fig:SGLM_thevol}. The inclinations of the rings for 
this model are seen to remain almost unchanged during the 
simulation. This is an effect of the low mass adopted for this 
model, and the corresponding long precession time. In order for this model 
to show a prominent breaking, the disc
would have to evolve for about $3.3 \times 10^4$ orbital periods 
($\approx 8 \times 10^6$ yr), which is longer than the age of the stars. 

\begin{figure}
\begin{centering}
\includegraphics[width=8.5cm, height=6cm]{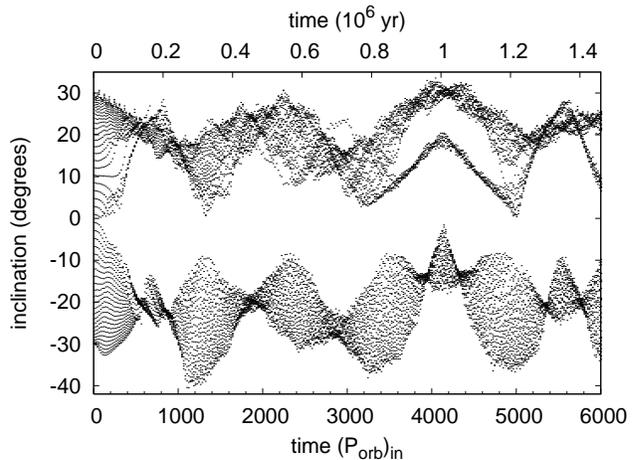}
\caption{Time evolution of the inclinations for model $\rm CLSG\_HM$. 
The disc breaks into pieces on time scales comparable to those for 
model $\rm SG\_HM$. The long time behavior of these two models differ in 
that when the cluster is present, the oscillations in inclination are less 
well behaved.}
\label{fig:CLSGHM_thevol}
\end{centering}
\end{figure}

In Fig. \ref{fig:CLSGHM_thevol} we show the warp evolution of model 
$\rm CLSG\_HM$ which considers the effect of both the self-gravity and 
the star cluster. We have seen in the previous subsection that for this 
mass fraction, it is the self-gravity that mainly dominates the 
precession evolution of the disc. The situation is therefore similar 
to that in Fig. \ref{fig:SGHM_thevol} when we consider 
the warp evolution. We see that model $\rm CLSG\_HM$ also 
breaks into pieces on time scales comparable to those for 
model $\rm SG\_HM$, attaining similar separation angles. 
More specifically, model $\rm CLSG\_HM$ reaches a maximum separation of 
$40^{\circ}$ at about 600 orbital periods. At 1300 orbital periods, 
the mutual inclination between the broken parts drops to $10^{\circ}$. This is 
twice the value attained in model $\rm SG\_HM$. However, the 
major difference between the two models shows itself in the long term 
evolution, the cluster model exhibiting a less well behaved 
oscillation pattern. 

\begin{figure}
\begin{centering}
\includegraphics[width=8.5cm, height=6cm]{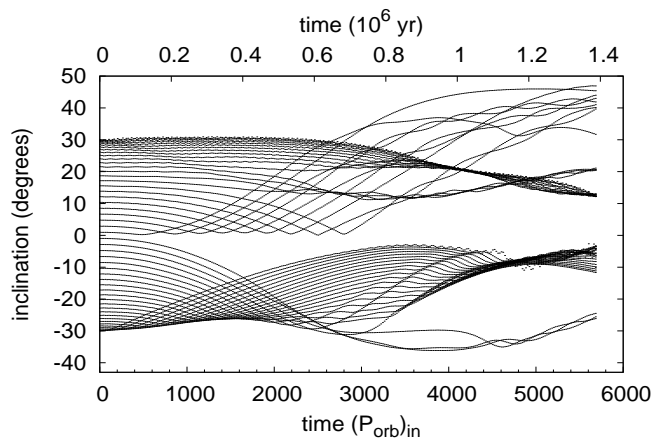}
\caption{Time evolution of the inclinations for model 
$\rm CLSG\_IM$. The self-gravity of the disc can not hold the rings 
together, and the disc breaks up into 
several fragments under the influence of the star cluster.}
\label{fig:CLSGIM_thevol}
\end{centering}
\end{figure}

When the mass fraction is decreased, the effect of the cluster torques 
become more prominent. This can be seen in Fig. \ref{fig:CLSGIM_thevol}. 
After about 500 orbital periods the ring inclinations start to 
deviate from their original values.

Compared to the other models in which the disc breaks up, the 
evolution  for this model is more disordered. There is no 
well defined pattern left in the warp profile once the 
disc breaks. After 2500 orbital periods, 
the positively inclined rings are pushed to 
inclinations of more than $30^{\circ}$, 
reaching a maximum of about $50^{\circ}$ at the end of the simulation. 
At this stage, only some of the inner rings exhibit a disc-like structure 
with a warp of about $10^{\circ}$. The original disc as a whole 
has dissolved due to the differential precession induced by the cluster, 
and the self-gravity of the disc is not able to 
hold the disc intact.

\begin{figure}
\begin{centering}
\includegraphics[width=8.5cm, height=6cm]{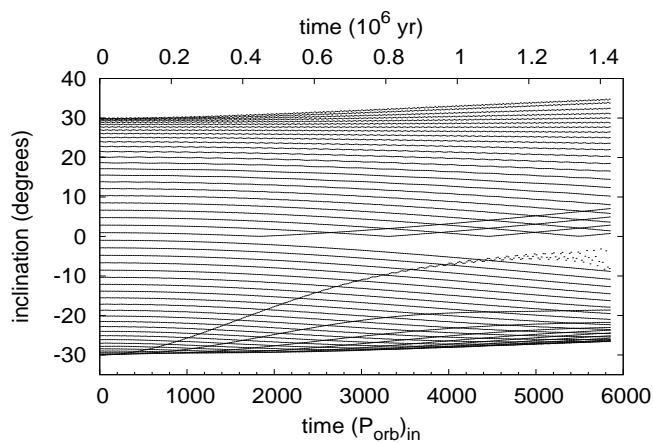}
\caption{Time evolution of the inclinations for model $\rm CLSG\_LM$. 
The inclinations do not change substantially while the rings 
precess apart azimuthally (see Fig. \ref{fig:CLSG_radphi}), 
because the mass fraction adopted is low for this model.}
\label{fig:CLSGLM_thevol}
\end{centering}
\end{figure}

In Fig. \ref{fig:CLSGLM_thevol} we show the evolution of 
inclinations for model $\rm CLSG\_LM$. For this model, we see 
that the inclinations are altered by only a few degrees, which at 
the end of the simulation leave the overall warp shape almost 
unchanged. This is because for this model the torques are 
dominated by the star cluster, leading to nearly free differential 
precession of the rings at constant inclination, causing the rings 
to spread in azimuth by about one full rotation 
(see Fig. \ref{fig:CLSG_radphi}).

\begin{figure*}$ \left. \right. $
\includegraphics{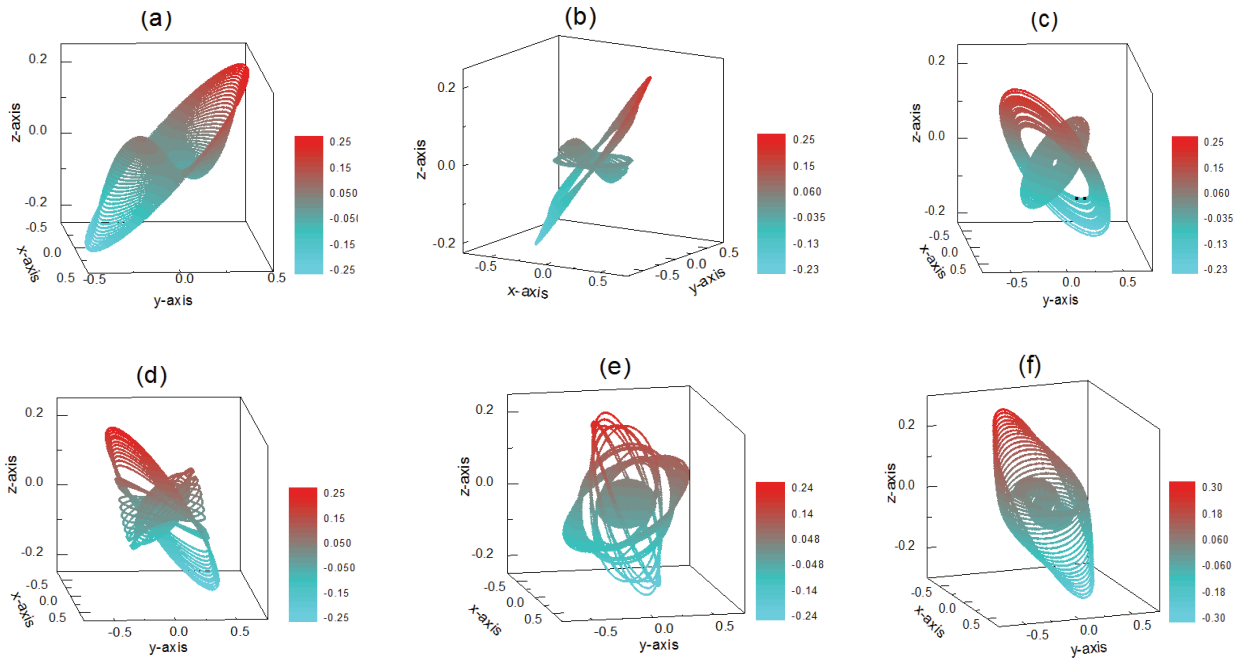}
\vspace{9.5cm}
\caption{3-dimensional visualizations of the warped disc models at 
different stages of their evolution. The color code displays the height 
($z$-coordinate) of the disc. Different viewing angles are chosen to highlight 
the various structures present in the warp shapes. The figures correspond to: 
the initial configuration common to all the models (a), 
model $\rm SG\_HM$ at $\rm \hat{t}=340$ (b), 
model $\rm SG\_IM$ at $\rm \hat{t}=8900$ (c), 
model $\rm CLSG\_IM$ at $\rm \hat{t}=2500$ (d),
model $\rm CLSG\_IM$ at $\rm \hat{t}=5680$ (e), and  
model $\rm CLSG\_LM$ at $\rm \hat{t}=5680$ (f) 
respectively where $\rm \hat{t}=t/ P_{orb_{\rm in}}$.}
\label{fig:3dplots}
\end{figure*}

In the last part of this section, we show in Fig. \ref{fig:3dplots} 
several examples of the 3-dimensional visualizations of our models at 
different stages of their evolution. The color code displays the height 
($z$-coordinate) of the disc. Different viewing angles are chosen to highlight 
the various structures present in the warp shapes. The figures correspond to: 
the initial configuration common to all the models (a), model 
$\rm SG\_HM$ at $\rm \hat{t}=340$ (b), model $\rm SG\_IM$ at $\rm \hat{t}=8900$ (c), 
model $\rm CLSG\_IM$ at $\rm \hat{t}=2500$ (d), model 
$\rm CLSG\_IM$ at $\rm \hat{t}=5680$ (e), and  model 
$\rm CLSG\_LM$ at $\rm \hat{t}=5680$ (f) respectively where $\rm \hat{t}=t/ 
P_{orb_{\rm in}}$. The broken shapes of the discs after some evolution for models $\rm SG\_HM$ and $\rm SG\_IM$ are 
apparent in figures (15b) and (15c). Fig. (15e) depicts clearly how the shape of a 
disrupted disc with parameters identical to those for model $\rm CLSG\_IM$
would look like. In the last panel in Fig. (15f), we see that the inner parts of 
the disc precessed more compared to the outer parts, while the inclination 
of the disc has only slightly changed compared to its initial 
state (see figures (\ref{fig:CLSG_radphi}) and (\ref{fig:CLSGLM_thevol})). 

\subsection{Comparison With the Observations}
\label{sec:observe}
In this section, we compare our time evolution 
models with the observations of the GC discs. In doing 
so, we will make use of the results of \citet{bartko09, bartko010}. 

After application of proper selection criteria, the sample 
includes a total of 136 stars on clockwise and counterclockwise 
rotating orbits. These stars have measured ($x,y$) position, 
while the line-of-sight distance $z$ is unknown. 
For the analysis, $z$ is therefore treated with a Monte-Carlo simulation.
The respective space velocities $v_x$, $v_y$, and $v_z$ are also given 
in \citet{bartko09}.

The number density of stars in the discs, $\Sigma_{\rm nr}(r)$, scales 
as $1/r^{1.4}$ \citep{bartko010}. Here $r$ is the 3-dimensional distance to SgrA*, 
corresponding to the radius $r$ of our rings.
The total number of stars on a ring is then 
$2 \pi r \Delta r\Sigma_{\rm nr}(r)$. If we assume that the total number of stars 
in the disc is $N_{\rm t}$ we can write
\begin{equation}
\displaystyle\sum_1^{n} 2 \pi r \Delta r \Sigma_{\rm nr}(r)=N_{\rm t},
\end{equation}
where $n$ is the number of rings making up the disc. 
We write $\Sigma_{\rm nr}(r)=\Sigma_0 (r/r_{\rm in})^{-1.4}$. 
The normalization $\Sigma_0$, i.e the number of stars on the innermost ring, is 
then obtained by writing
\begin{equation}
\Sigma_0=\frac{N_t}{2 \pi  \Delta r \ r_{\rm in}^{1.4}  \displaystyle\sum_1^{n}  \frac{1}{r^{0.4}}}.
\end{equation}
The coordinates of a star on a ring are given by
\begin{eqnarray}
x & = & r (\cos \phi \cos \psi - \cos \theta \sin \phi \sin \psi), \nonumber \\
y & = & r (\sin \phi \cos \psi + \cos \theta \cos \phi \sin \psi), \nonumber \\
z & = & r \sin \theta \sin \psi.
\label{eq:coord}
\end{eqnarray}

In order to construct the phase space distributions of the stars populating 
the discs,  we write the components of the stellar velocities
\begin{eqnarray}
v_x&=&r(\dot{\theta} \sin \theta \sin \psi \sin \phi -\dot{\psi} \sin \psi \cos \phi -\dot{\phi} 
\sin \phi \cos \psi  \nonumber \\
&-&\dot{\phi} \cos \theta \cos \phi \sin \psi  - \dot{\psi} \cos \theta \sin \phi \cos  \psi), 
\nonumber \\
v_y &= &r (\dot{\phi} \cos \phi \cos \psi - \dot{\phi} \cos \theta \sin \phi \sin \psi \nonumber \\
&+&\dot{\psi} \cos \theta  \cos \phi \cos \psi -  \dot{\theta} \sin \theta \sin \psi \cos \phi 
-\dot{\psi} \sin \psi \sin \phi ), \nonumber \\
v_z &=& r(\dot{\theta} \sin \psi \cos \theta + \dot{\psi} \sin \theta \cos \psi).
\label{eq:vel}
\end{eqnarray}
Since we know the values of ($\dot{\phi}, \dot{\theta} , \dot{\psi}, \phi, 
\theta$) from our time integration, we can generate a random sample of stars 
on these rings, by dicing $\psi$. We can then construct the phase space 
distribution of stars at each instant of time $t$ using 
equations (\ref{eq:coord}) and (\ref{eq:vel}) . 

The GC contains two warped and almost orthogonal 
discs of young stars which extend over a similar range of projected 
radii \citep{bartko010}. The CW disc is observed to extend between 
projected radii of approximately 0."8 and 15", while the CCW disc 
is seen between projected radii of about 3."5 and 15". This makes 
it unlikely that the two stellar discs derive from a single 
progenitor disc, but rather suggests that they might be tracers of two 
separate accretion episodes. In the following, we will concentrate 
in our model comparison on the CW disc, which is better defined due 
to its larger number of constituent stars.

We have shown in the previous section that massive discs break 
into pieces. Although the broken discs may precess as coherent separate discs 
under certain circumstances, their degree of warping is highly reduced 
compared to the original disc. 
The observational analysis presented in \cite{bartko09} can discriminate 
between a disc precessing as a single unit and a broken disc, 
such that the former produces a warp profile that changes gradually 
between the inner and the outer radii, while the latter would 
result in a jump at a certain radius. 

A useful way of comparing the models with the observations 
is to construct the components of angular momenta 
and see how they change with projected 
radius. The component of angular momentum along the axis 
pointing towards the observer will then give information on the 
twist of the disc, while the component projected onto the plane 
of the sky will give the degree of the warp 
(for further details on the definitions of the 
angles see \cite{bartko09}).

\begin{figure*} $ \left. \right. $
\includegraphics{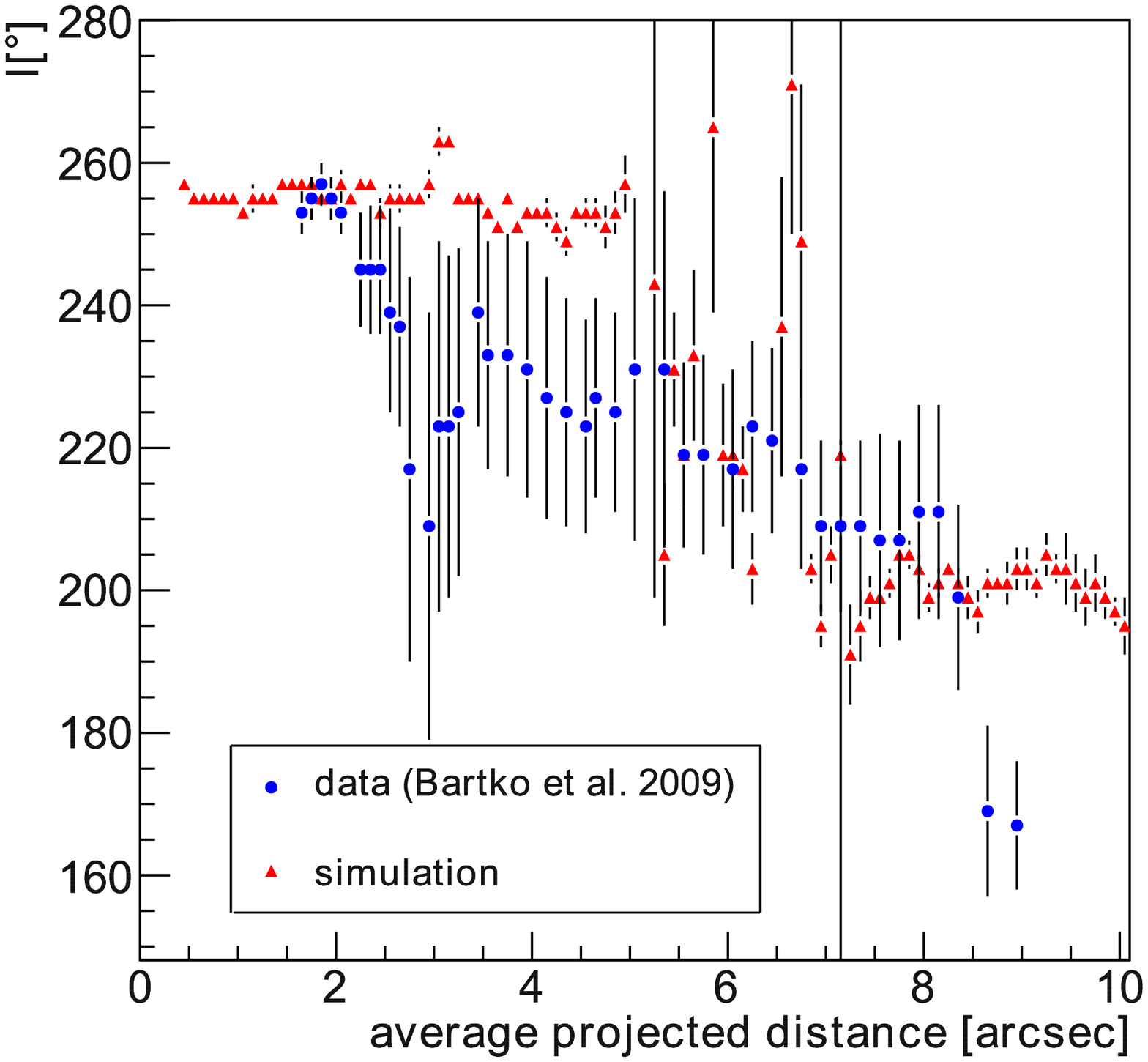}
\includegraphics{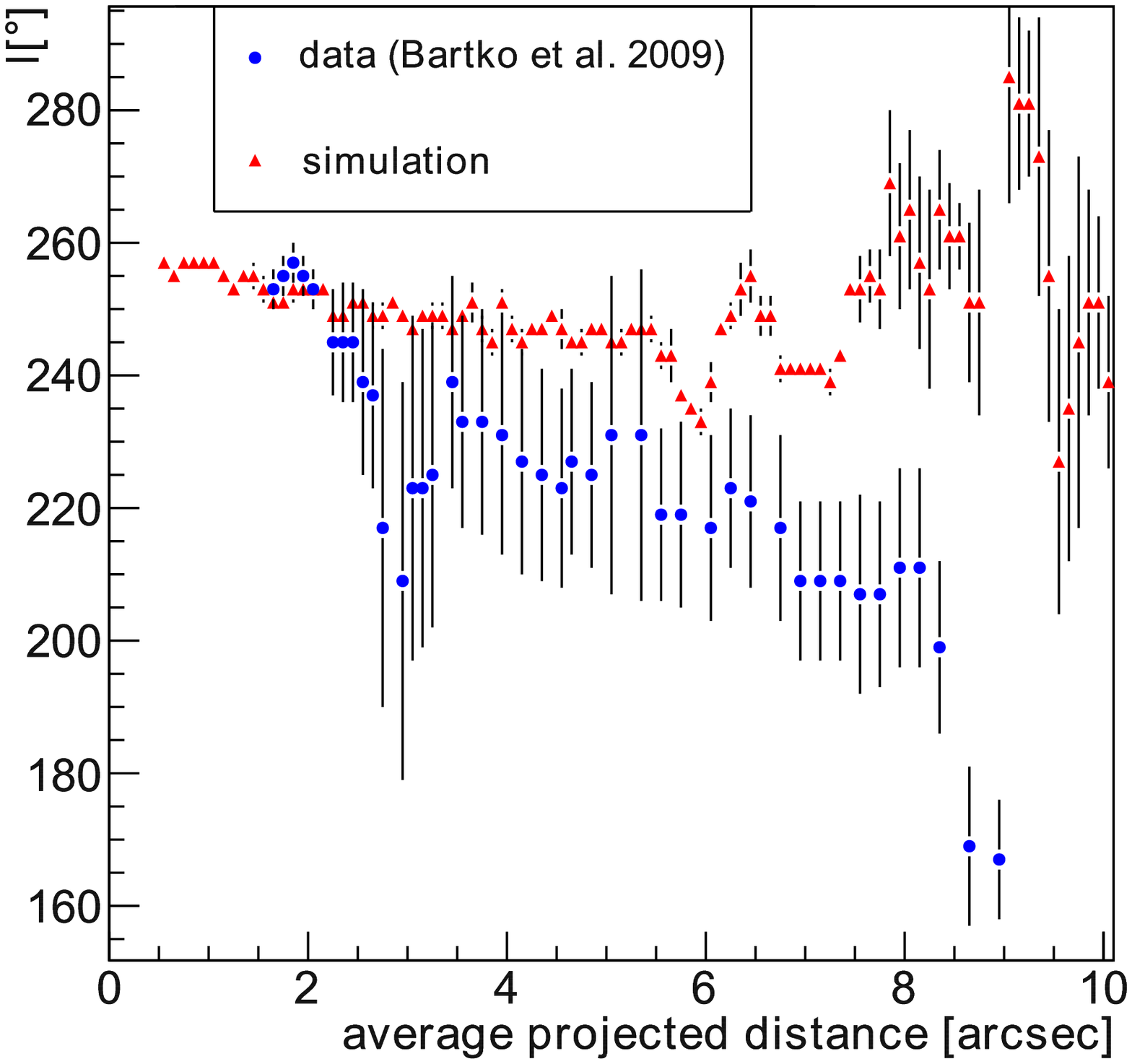}   
\includegraphics{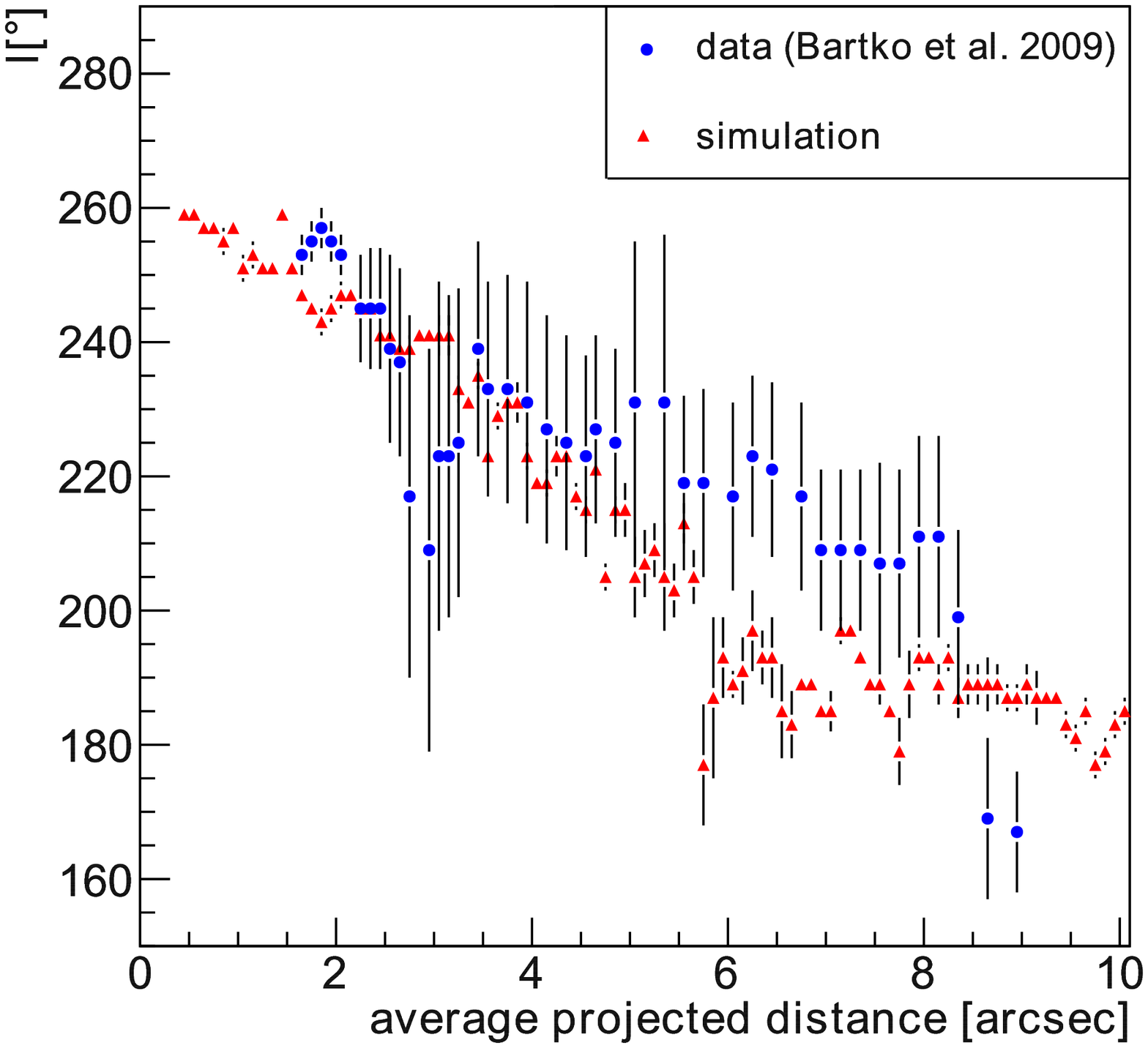}    
\includegraphics{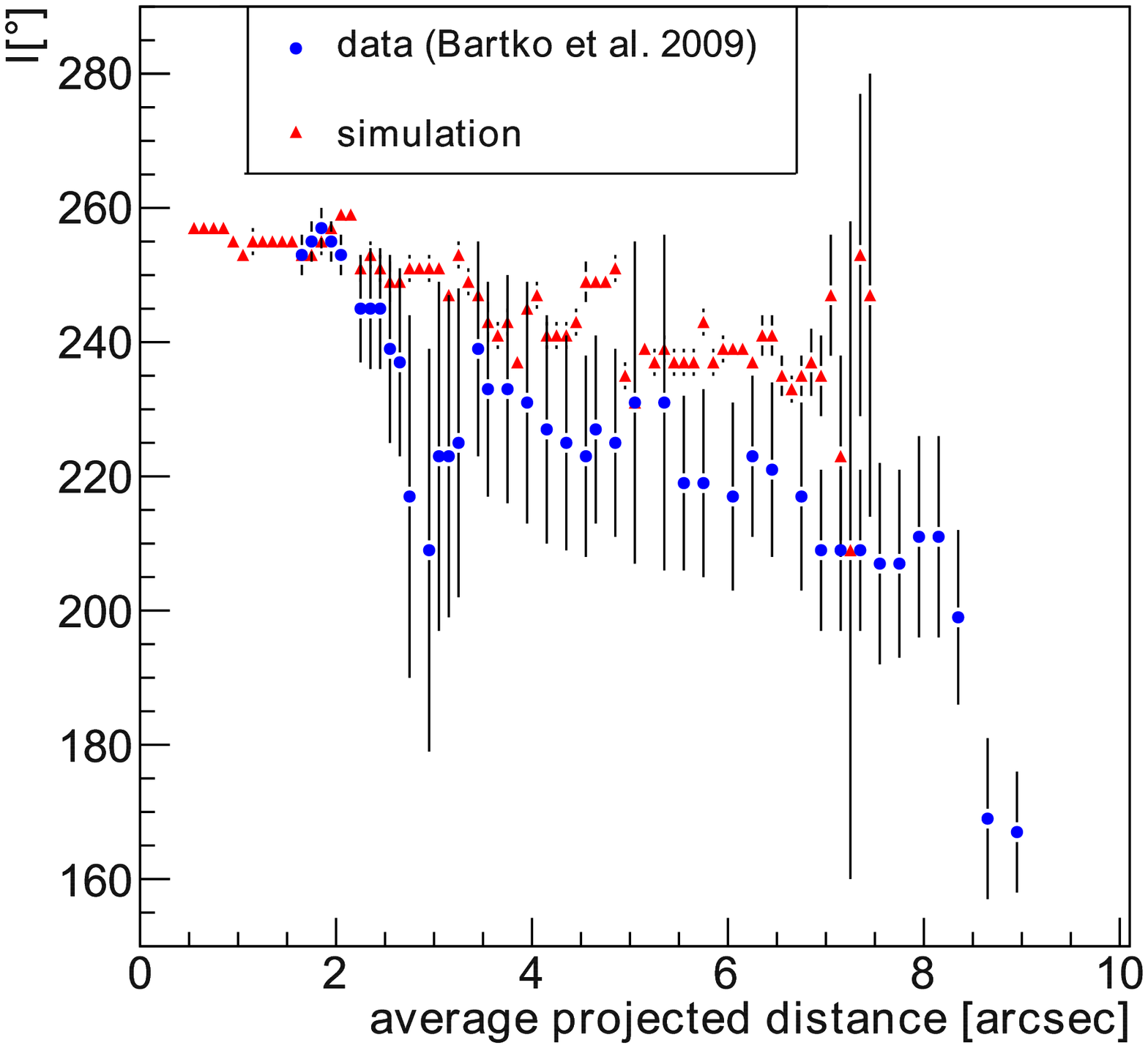}  
\includegraphics{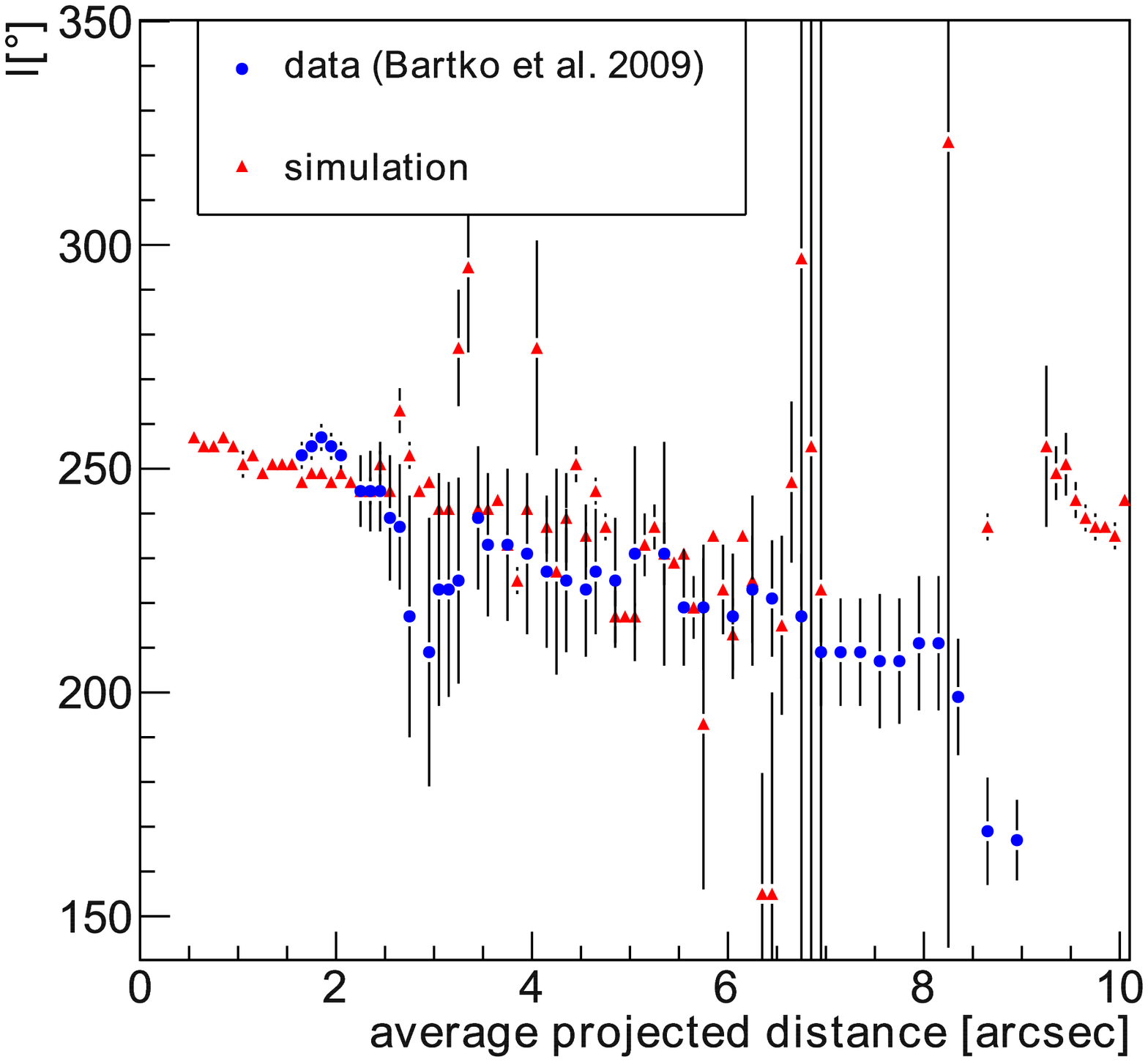}    
\includegraphics{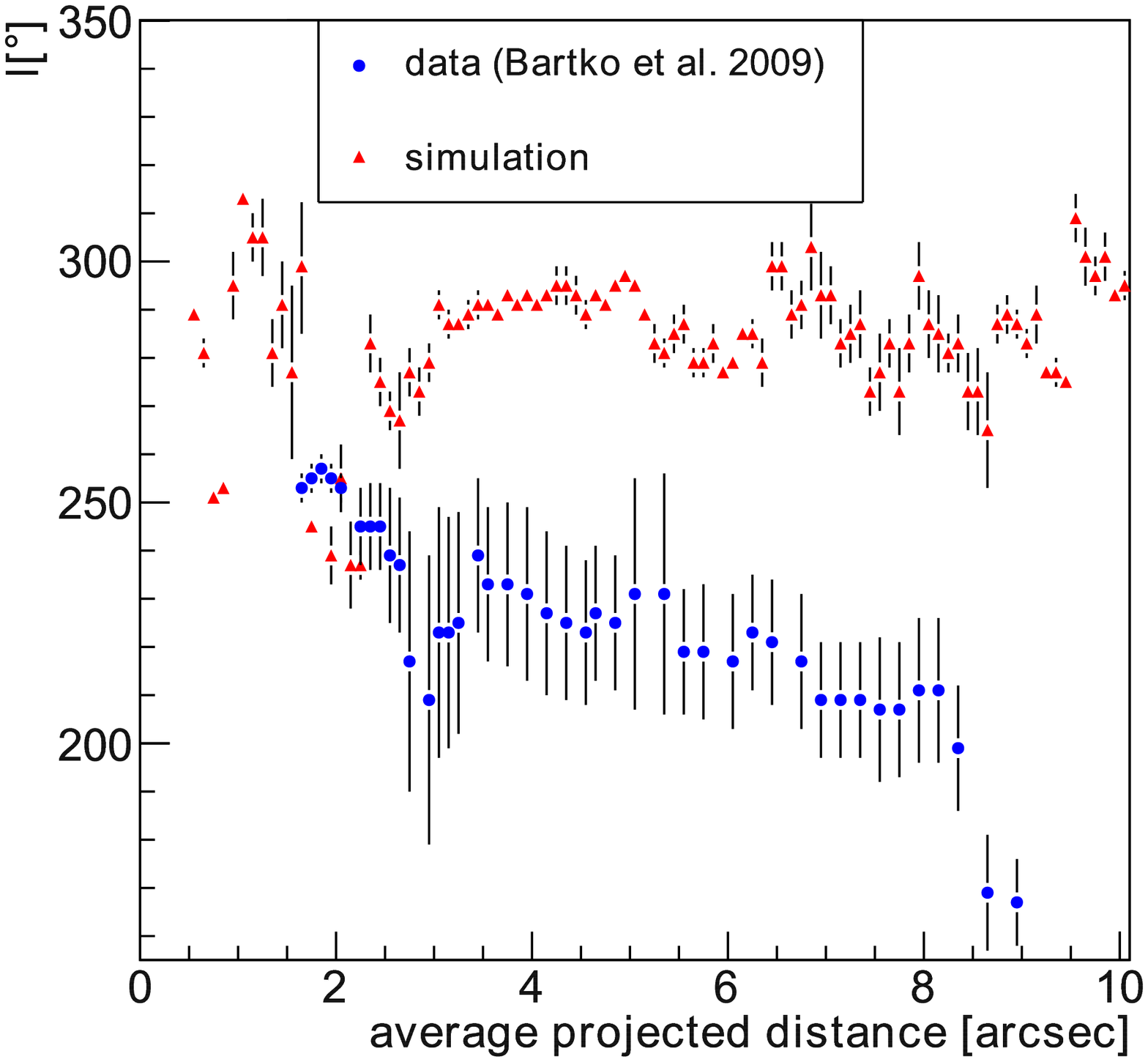}
\vspace{9.8cm}
\caption{The warp profiles of the model discs (red triangles) and of the CW 
disc (blue dots). The figures on the top show the 
self-gravity-only models, and the ones at the bottom show the 
models including the effects of the star cluster. The mass fraction 
of the simulated discs decreases going from left to right. The lowest 
mass model for the case of self-gravity-only gives the best agreement with the data.}
\label{fig:phase}
\end{figure*}

Fig. \ref{fig:phase} shows the component of 
the local average angular momentum direction of the model 
discs (red triangles), and of the CW disc as 
published in \cite{bartko09} (blue dots) on the plane of 
the sky as a function of average projected distance from 
the centre (l=$\phi_{J}(R)$ in the notation of \cite{bartko09} 
where $J$ denotes the angular momentum per mass). The data from the 
simulations is rotated successively such that the orientations 
of the model and the observed discs match at the inner edge, and 
the warp in the simulations fit the observed warp best. The points 
referring to the simulations are 
chosen at certain times during the late stages 
of the evolution of the discs. 
The figures on the top are for the self-gravity-only models, and the 
ones at the bottom are for the 
models including the effects of the star cluster. The mass fraction of the 
simulated discs decreases going from left to right, therefore the models 
should be read as $\rm HM$, $\rm IM$, and $\rm LM$ going in this direction.  

We see that a good agreement with the observations of the CW disc 
is achieved for our model $\rm SG\_LM$ in that the inclination 
(warping of the disc) changes gradually between the 
inner and the outer edges of the disc, 
and that the amplitude of the warp is successfully reproduced. For the 
self-gravity-only models, higher mass fractions lead to flattening of 
the inner parts of the disc, hence can be excluded 
to be representative of the data. On the other hand, when the torques from the star cluster 
are taken into account, none of these models 
seem to match the data well.

\section{Discussion}
\label{sec:discussion}
\subsection{Dependence on the Initial Conditions}
\begin{figure}
\begin{centering}
\includegraphics[width=8.5cm, height=6cm]{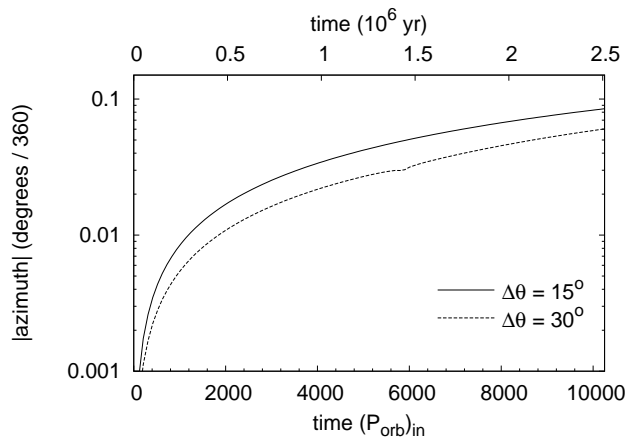}
\caption{Time evolution of the azimuthal angles for models LI (solid line), 
and $\rm SG\_LM$ (dashed line). Model LI precesses 1.3 times faster than 
model $\rm SG\_LM$ due to its decreased overall inclination.}
\label{fig:dep_init_conds_phis}
\end{centering}
\end{figure}

In the previous section we argued that a good agreement with 
the observations of the CW disc is achieved for our model $\rm SG\_LM$. 
To test the importance of the choice of the initial warp configuration we performed 
a run, model LI, where the initial inclinations, i.e. $\Delta \theta$, span a 
range $-15^{\circ}$ to $15^{\circ}$, and the 
other set of parameters are identical to those of model $\rm SG\_LM$. 
Fig. \ref{fig:dep_init_conds_phis} shows the evolution of the azimuthal angles 
for models LI (solid line), and $\rm SG\_LM$ (dashed line). We see that when the disc 
is less inclined it precesses faster, by a factor of about 1.3 for the parameters 
chosen. An analytic expression showing the inverse proportionality of the precession 
rate to the disc inclination for steadily precessing discs can be found in US09. 

\begin{figure}
\begin{centering}
\includegraphics[width=8.2cm, height=6cm]{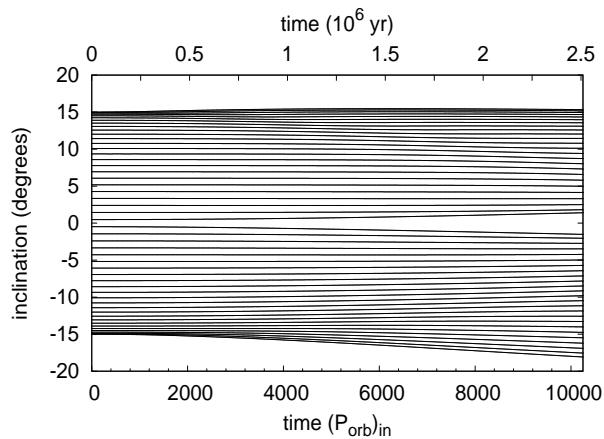}
\caption{Time evolution of the ring inclinations for model LI. The ring 
inclinations change only slightly throughout the simulation, leading to a 
final warp of about $32^{\circ}$.}
\label{fig:dep_init_conds_th}
\end{centering}
\end{figure}

We show in Fig. \ref{fig:dep_init_conds_th} the evolution of the 
ring inclinations for model LI. We see that the splitting of the disc 
into parts, although not much pronounced, occurs on a shorter time as 
compared to model $\rm SG\_LM$ (the inclinations in model $\rm SG\_LM$ had 
remained unaltered during a similar time span). Even if there are slight 
excursions of the ring inclinations from their initial values, the overall 
warp shape does not change considerably throughout the simulation for model 
LI either. Hence we conclude that in order for the low mass disc to evolve towards 
the observed orbital configuration, its initial inclination determined by the 
warping mechanisms has to be close to the observed inclination. 

\subsection{Comparison to Linear Evolution}
In US09 it is shown that linear theory 
over estimates the torques acting on the edges of the disc, leading 
to a slightly less inclined warp configuration at its maximum possible 
value (within the validity of the linear theory). 
Here, we would like to briefly discuss 
the evolution of the disc in linear theory, where the 
linearization is done as in US09. In order to reduce computational 
time we have considered one purely self-gravitating model adopting 
a mass fraction of $M_{\rm d}/M_{\rm bh}=0.134$. We have run 
the simulation for about $\rm 1250 P_{orb_{in}}$, which corresponds 
to half a precession time for the non-linear model for this mass fraction. 
Since the evolution 
of the disc scales approximately linearly in mass, the comparison 
which we will present should equally apply to discs with other mass fractions. 

\begin{figure}
\begin{centering}
\includegraphics[width=8.2cm, height=6cm]{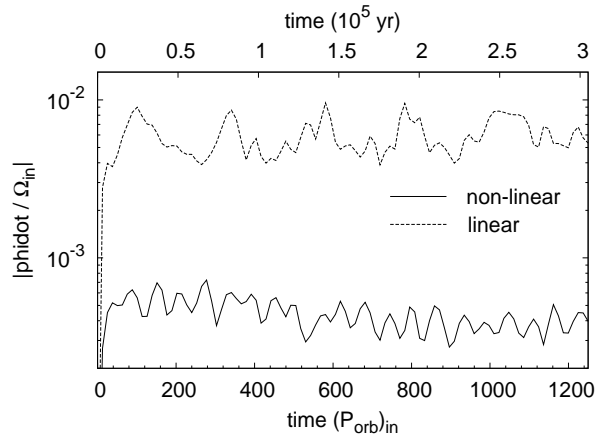}
\caption{Average precession frequencies attained during the simulations 
for a mass fraction of $M_{\rm d}/M_{\rm bh}=0.134$. The solid line is 
for model $\rm SG\_HM$ calculated in the non-linear torques regime, and 
the dashed line is for the linear model. The disc precesses with a much 
higher rate in the linear regime.}
\label{fig:linear_phidots}
\end{centering}
\end{figure}

In Fig. \ref{fig:linear_phidots} we show the averaged precession 
frequencies of the non-linear (model $\rm SG\_HM$) (solid line), 
and the linear models (dashed line). We see that in linear theory the 
model acquires an average 
precession frequency of 
$\dot{\phi} / \Omega_{\rm in} \approx -6 \times 10^{-3}$, 
corresponding to a precession time scale of $4 \times 10^4$ yr. Hence the 
model disc in linear theory precesses 15 times faster than it should. Turned 
around, the system behaves as if it were 15 times more massive (for the mass 
fraction considered here, this would unphysically indicate a disc twice 
more massive than the central black hole). 

\subsection{Limitations}
\subsubsection{Eccentricity of the Disc}
In our simulations, we used a simple circular ring model although 
the observations point to an eccentricity of the 
orbits with $e \sim 0.35$  for the CW disc and close 
to zero eccentricity for the CCW disc \citep{paumard06, bartko09, bartko010}. 
The change of eccentricity (as well as inclination) in 
an initially circular disc might be driven by the interaction 
of the disc stars with the surrounding cusp \citep{perets08, loeckmann09}. 
Following the time evolution of single and initially mutually inclined two 
discs, \cite{Loeckmann09a} showed that such interactions can cause the inner 
parts of the disc, i.e. $r < 0.3$ pc, to gain quite high eccentricities on a typical 
time scale of 5 Myr. Neglecting the effects of the disc's 
self-gravity \cite{madigan09} showed that an initially eccentric disc embedded in a 
stellar cusp around a massive black hole is subject to an instability which leads to 
the growth of eccentricities on about a precession time ($\sim$ 0.6-0.7 Myr at 0.05 pc).  
It is not known whether taking into account the self-gravity 
of the disc would modify the results of these calculations 
strongly. 

\subsubsection{Viscous Evolution of The Disc}
In proposing our scenario, we made a strong assumption that 
there had been an accretion disc around SgrA* a few million years ago. 
While we have argued for the plausibility of disc warping 
prior to star formation, our simulations were performed following 
only the gravitational evolution of the disc. The next step 
forward would be to consider the viscous evolution of the disc under 
the non-linear self-gravity torques, as well as the torques from 
the old star cluster. As we will briefly discuss in the next 
subsection, viscous discs in various physical environments are 
also prone to be broken or disrupted by differential precession.  

\subsection{Vector Resonant Relaxation as a Warping Mechanism}
In our scenario, we considered two mechanisms, the radiation 
instability, and the Bardeen-Petterson effect to explain the 
warped gaseous disc origin of the GC discs. 
Without making a detailed numerical analysis of these 
mechanisms, we argued for their possibility based on 
simple arguments such as the critical radii and time scales 
for warping. It would be interesting to explore other 
warping mechanisms than the two discussed here. 
Recently, \cite{bregman09} showed that vector resonant 
relaxation (VRR) resulting from the interaction of the disc 
with the surrounding old star cluster might explain the warp 
in the maser disc of NGC4258. When scaled to the parameters of the 
GC region, VRR serves as a viable mechanism for orbits within 
the central $\sim 0.2$ pc \citep{kocsis011}. Beyond 
this radius, the VRR time scale exceeds the life time of the disc 
stars. However, it would still be interesting to see how the disc 
responds to the stellar cusp when the simulations are carried 
out in the non-linear regime.

\subsection{Broken/Disrupted Warped Discs}
We have shown examples of purely gravitating warped disc 
simulations where the discs with a certain mass fraction break 
into pieces. This is an effect of the faster precession of some parts 
of the disc, and can lead to a complete disruption of the disc if 
the broken parts can not be held together by internal torques. 

Broken or disrupted discs have been reported before mainly in the context 
of viscous discs. Accretion discs warped by tidal effects in 
misaligned binaries can be disrupted due to differential 
precession when the disc aspect 
ratio $h=H/r \sim 0.01$, where $H$ is the disc height \citep{larwood96, fragner010}. 
When the warp propagation through the disc is diffusive rather 
than wave-like, isothermal discs with small viscosity exhibit a steepening in 
their warp profiles where the disc breaks into two parts \citep{ogilvie06,lodato010}.

\section{Summary and Conclusions}
\label{sec:summary}
We have proposed a new scenario for the formation of the warped stellar 
discs of young stars at the Galactic Centre (GC). In contrast to the work 
published so far, our study is based on the assumption that the star 
formation at the GC took place in an accretion disc but 
was delayed until the accretion disc around SgrA* was warped. To test the 
plausibility of this idea we addressed the issue of accretion disc 
warping considering two mechanisms, the Pringle (radiation) 
instability, and the Bardeen-Petterson effect. From simple arguments 
we showed that warping of the GC disc is plausible if the 
star formation efficiency is high, $\epsilon_{\rm SF} \lesssim 1$, and 
the viscosity parameter $\alpha \sim 0.1$. 

Our simulations are based on the idea that, some time after 
the disc became warped, the activity at the GC subsided.  
Rapid cooling of the disc led to fragmentation and star formation,
while the rest of the gas was lost. Subsequently, the disk was
subject only to gravitational forces causing it to precess. In 
addition to the 
torques from the self-gravity of the disc, the torques induced by a 
surrounding non-spherical star cluster were also taken into account 
in our models. We examined the long term temporal behavior of 
the warped disc models with different disc-to-black 
hole mass ratios (mass fractions), 
and investigated the interplay between the self-gravity and the 
cluster torques. We also made a comparison of our models with 
the observations of the warped GC discs. 
Our findings can be summarized as follows: 

$i)$
When the disc mass fraction is large, 
$M_{\rm d}/M_{\rm bh} \approx 0.1$, the disc 
breaks into distinct pieces which then precess 
independently. In this case, the self-gravity torques 
dominate over the torques from the star cluster, for cluster 
parameters representative for the GC.
The warp profiles of these models evolve in an oscillatory manner 
which is very well behaved for the self-gravity model. The maximum 
mutual inclinations attained (about $40^{\circ}$), and the time 
scales by which the disc acquires this broken configuration are 
comparable in both models, despite a slightly 
faster precession of the cluster model. 

$ii)$ Moderate mass discs with $M_{\rm d}/M_{\rm bh} \approx 0.01$ 
evolve in two distinct ways, depending on the absence 
or presence of the torques from the star cluster. In the 
former case, the evolution of the model mimics 
that of its higher mass counterpart. 
Both the precession and the break-up of the disc 
occur on time scales which are almost exactly 10 times longer 
than for the high mass model. The model 
including the effect of the star cluster for this mass fraction exhibits 
a more disordered evolution. The disc breaks into parts more slowly, 
but once it arrives in a broken configuration, 
its pieces cannot stay intact due to the decreased self-gravity, 
and the disc as a whole dissolves.
 
$iii)$ Low mass discs for which $M_{\rm d}/M_{\rm bh} \approx 0.001$ 
precess on time scales of order a few $10^7 $ yr when the disc is 
subject only to its own self-gravity, while the presence of the 
star cluster fastens the precession of the disc by about 
a factor of fourty for the estimated quadrupole moment of 
the star cluster. The overall inclination of the disc is 
not grossly altered during the simulations for this mass 
fraction. When only the self-gravity of the disc is considered, 
the disc precesses as a single body throughout its evolution, while 
inclusion of the torques from the star cluster cause the 
rings to precess apart. 

$iv)$ None of our models lead to a configuration where parts of a broken 
disc would counter-rotate on the plane of the sky, i.e. give rise 
to two mutually counterrotating discs. A comparison of our models with the
better-defined clockwise rotating disc shows that the lowest mass 
self-gravity only model matches the data best. 
In this case in order for the model to mimic the observed orbital 
configuration of the stellar disc, 
its initial inclination determined by the gas disc warping mechanism 
has to be close to the observed one. On the other hand when the star cluster torques 
are taken into account, the overall shape of the 
warp can not be reproduced. The disagreement of the model including 
the star cluster might mean either that the old star cluster at the Galactic 
Centre is almost spherical so that it exerts no torque on the discs, or that 
the mass of the cluster within the radius of the disc is less than the 
value assumed here based on the most 
recent observations. Future observations of the old star cluster will 
therefore help to better understand the behavior of the stellar discs. 

\section*{Acknowledgments}
A.U.S would like to thank M. Arnaboldi for sharing ideas on warped discs 
in a general context, and R. Sunyaev for pointing to useful 
literature in the early stages of this study. She also acknowledges 
the support of staff at CfA, Harvard University where part of this work was 
completed. This work is partially supported by the Scientific Research Projects 
Coordination Unit of Istanbul University under project number UDP-16581.

\bibliography{GCdiscs}

\label{lastpage}

\end{document}